\documentclass[twocolumn]{aastex631}

\usepackage{amsmath}
\usepackage{listings}
\usepackage{tabularx}
\usepackage{longtable}
\lstdefinestyle{sbtext}{
    basicstyle=\ttfamily\scriptsize,
    breaklines=true,
    columns=fullflexible,
    keepspaces=true,
    frame=single
}
\usepackage{xcolor}

\usepackage{array}
\usepackage{booktabs}
\usepackage{comment}

\begin{document}

\title{Toward Autonomous Radio Follow-up of Multi-messenger Transients with \texttt{RADAR}: \\From Alert Parsing to Inference and Observation Scheduling}

\author[0000-0003-4523-4807]{Mihael Hategan-Marandiuc}
\affiliation{Department of Computer Science, The University of Chicago, Chicago, Illinois 60637, USA}
\affiliation{Data Science and Learning Division, Argonne National Laboratory, Lemont, Illinois 60439, USA}

\author[0009-0007-1842-7028]{Tanner~O'Dwyer}
\affiliation{William H. Miller III Department of Physics and Astronomy, Johns Hopkins University, Baltimore, Maryland 21218, USA}

\author[0000-0001-8104-3536]{Alessandra Corsi}
\affiliation{William H. Miller III Department of Physics and Astronomy, Johns Hopkins University, Baltimore, Maryland 21218, USA}

\author[0000-0002-9682-3604]{Eliu Huerta}
\affiliation{Data Science and Learning Division,
Argonne National Laboratory, Lemont, Illinois 60439, USA}
\affiliation{Department of Computer Science, The University of Chicago, Chicago, Illinois 60637, USA}
\affiliation{Department of
Physics, University of Illinois Urbana-Champaign,
Urbana, Illinois 61801, USA}
\affiliation{Department of
Astronomy, University of Illinois Urbana-Champaign,
Urbana, Illinois 61801, USA}

\author[0000-0002-0160-2519]{Zilinghan Li}
\affiliation{Data Science and Learning Division,
Argonne National Laboratory, Lemont, Illinois 60439, USA}

\author[0000-0003-2129-5269]{Ian T. Foster}
\affiliation{Data Science and Learning Division,
Argonne National Laboratory, Lemont, Illinois 60439, USA}
\affiliation{Department of Computer Science, The University of Chicago, Chicago, Illinois 60637, USA}

\author[0000-0002-7370-4805]{Kyle Chard}
\affiliation{Department of Computer Science, The University of Chicago, Chicago, Illinois 60637, USA}
\affiliation{Data Science and Learning Division,
Argonne National Laboratory, Lemont, Illinois 60439, USA}

\author[0000-0002-6781-7432]{Ryan Chard}
\affiliation{Data Science and Learning Division,
Argonne National Laboratory, Lemont, Illinois 60439, USA}

\author[0009-0004-4830-3139]{Amal Gueroudji}
\affiliation{Mathematics and Computer Science Division,
Argonne National Laboratory, Lemont, Illinois 60439, USA}
 
\begin{abstract}
Multi-messenger astronomy (MMA), the joint study of cosmic sources through gravitational waves (GWs), electromagnetic (EM) radiation, neutrinos, and cosmic rays, is rapidly reshaping time-domain astrophysics. Realizing the promise of MMA will require coordinating heterogeneous observing resources and automating the chain from alert to analysis to follow-up. \texttt{RADAR} (\texttt{R}adio \texttt{A}fterglow \texttt{D}etection and \texttt{A}I-driven \texttt{R}esponse) is a federated, privacy-enhancing framework for the radio follow-up of GW events, previously validated on GW170817. Here, we extend it along three axes. First, we benchmark three large language models (LLMs; \texttt{GPT-5.5}, \texttt{Claude-Opus-4.7}, and \texttt{Gemini-3.5-Flash}) against the GW170817 radio light curve dataset. 
\texttt{GPT-5.5} attains the highest event-level $F_1$ score, the harmonic mean of precision and recall, at $0.893 \pm 0.010$, and the highest GCN-level recall, $0.794 \pm 0.013$, improving on previous \texttt{GPT-4.1} results 
by 16\% and 10\%, respectively, while \texttt{Claude-Opus-4.7} achieves the highest precision, $0.978 \pm 0.014$.
Second, we introduce concurrent likelihood evaluation, which speeds up the MCMC computation by a factor of $40\times$ over our previous results. Third, we present an LLM-driven scheme that converts natural-language observing requests into submission-ready scheduling blocks for the Karl G. Jansky Very Large Array. Together, these developments advance \texttt{RADAR} toward a scalable, largely autonomous system for GW radio follow up.

\end{abstract}

\keywords{Gravitational waves;  stars: neutron; radio continuum: general; Methods: data analysis}

\section{Introduction} 
\label{sec:intro}

The binary neutron star (BNS) merger GW170817 \citep{2017PhRvL.119p1101A} inaugurated the era of multi-messenger astronomy \citep{2019NatRP...1..585M}, with its near-simultaneous detection in gravitational waves (GWs) and across the electromagnetic (EM) spectrum \citep{2017ApJ...848L..13A}. This discovery confirmed BNS mergers as progenitors of short gamma-ray bursts (GRBs) and their relativistic jets \citep[e.g.,][]{2017ApJ...848L..12A,2017ApJ...848L..21A,2017Sci...358.1579H,2017ApJ...848L..20M,2017ApJ...848L..15S,2017Natur.551...71T,2018Natur.561..355M,2018Natur.554..207M,2019Sci...363..968G}, and as sites of heavy-element nucleosynthesis \citep[e.g.,][]{2017ApJ...848L..19C,2017ApJ...848L..17C,2017Sci...358.1570D,2017Sci...358.1565E,2017Natur.551...80K,2017Sci...358.1559K,2017LRR....20....3M,2017Sci...358.1574S,2017ApJ...848L..16S,2017ApJ...848L..24V,2017ApJ...851L..21V,2018PhRvL.120x1103L}.

A key lesson learned from GW170817 and subsequent observing runs of the LIGO--Virgo--KAGRA (LVK) network \citep{Abbott_2019_GWTC1,LIGOScientific:2020ibl,Abbott_2023_GWTC3,LVK:2025gwtc4,Abbott_2026_GWTC5_intro} is that the scientific return of multi-messenger campaigns depends critically on the timely coordination of many independent facilities and data analysis strategies \citep[e.g.,][]{2012SPIE.8448E..0QS,2012A&A...539A.124L,2018LRR....21....3A,2019ApJ...875..161A,2019PhRvD.100j3025C,2019PhRvD.100f3015G,2019NatRP...1..600H,2020ApJ...905L..25S,2021ApJ...910L..21M,2021PhRvD.104b3014M,2021PhRvD.103j3006W,2022NatPh..18..112G,2022ApJ...924...54P,2023ApJ...958L..43H,2018PhLB..778...64G,2024arXiv241104793A,2024APh...15502904A,2024CQGra..41h5012B,2024FrASS..1186748C,2024MLS&T...5d5030C,2024PhRvD.109d2008E,2024ApJ...963...98R,PhysRevD.97.044039,2024EPJWC.29504022V,2025LRR....28....2C,2025Natur.639...49D,2025PhRvD.111d2010M,2025arXiv251212513T,2025RSPTA.38340126N,2026ApJ...998....8S}. Before the true EM counterpart of a single GW event is identified, multiple observatories may need to follow up a range of candidate transients, and these observations must then be synthesized into a coherent physical picture. Collecting the data needed to discriminate among competing models and break parameter degeneracies before the transient fades is therefore critical. Yet observing resources are scarce, and their efficient use is essential, all the more so because EM counterparts at different wavelengths probe distinct ejecta components and evolve on vastly different timescales, ranging from seconds in gamma-rays to months or years in the radio. 

As GW detectors become more sensitive and detection rates increase \citep{2015CQGra..32b4001A,2015CQGra..32g4001L,2018LRR....21....3A,2019NatAs...3...35K,2021arXiv210909882E,2023arXiv230613745E,2025arXiv250803392K,2026JCAP...03..081A}, the central challenge will gradually shift: rather than focusing on identifying genuine EM transients among numerous candidates found in the large localization areas of a trickle of poorly localized GW events, the community will increasingly need to coordinate the follow-up of a large number of high-confidence, relatively well localized ($\mathcal{O}(10)$\,deg$^{2}$) detections \citep[e.g.,][and references therein]{2024FrASS..1101792C,2024FrASS..1186748C,2024CQGra..41x5001G,2024FrASS..1101785H}. However, the fundamental requirement will remain unchanged: the development of robust frameworks that enable rapid follow up and information sharing, distributed analysis, and automated decision-making across the multi-messenger community \citep[e.g.][and references therein]{2021NatAs...5.1062H, 2019NatRP...1..600H,2024arXiv240102063T,2025ApJS..280...71P,2020ApJ...894..127W}. 

In this context, the radio band offers a unique probe of the fast, non-thermal ejecta of compact binary mergers, but radio follow-up is uniquely demanding: sensitive interferometers see only tiny patches of the large GW localization regions, the emission can evolve over weeks to years \citep[e.g.,][and references therein]{Hallinan_2017,2024FrASS..1101792C,2026ApJ..1006..215M}, and coordination can be hampered by heterogeneous data-sharing policies (from immediate public release to long proprietary periods). In a prior paper, \cite{patel2025ai4mma_code}, we introduced \texttt{RADAR} (Radio Afterglow Detection and AI-driven Response) to meet these challenges through community information sharing, federated data analysis, and AI-assisted automation, while preserving each group's data rights and autonomy. \texttt{RADAR} comprises four linked modules: a GW module that runs AI models at distributed detector endpoints and sends only compact embeddings to a central server; a radio module that aggregates public ``General Coordinates Network'' (GCN) circulars and proprietary data, using a continuously running listener and an LLM-powered parser to extract the structured metadata needed to build radio light curves; and a multi-messenger module that combines GW and radio constraints via shared-parameter posteriors. These modules are connected by the \texttt{Octopus} event fabric~\citep{pan2024octopus}, 
a cloud-to-edge messaging layer enabling asynchronous, privacy-enhancing exchange of intermediate results across sites. 

\texttt{RADAR}'s defining feature is its federated approach to radio light-curve modeling: instead of centralizing proprietary observations, it coordinates fitting by exchanging model parameters and partial likelihoods, so raw data never leave their originating site while all sites contribute jointly to the inference. Built on \texttt{afterglowpy} \citep{2020ApJ...896..166R}, the framework supports both an exact distributed-likelihood scheme and a faster consensus (posterior-averaging) approximation, and allows progressive inference as new data arrive. The use of LLMs for structured information extraction from astronomical text reflects a broader trend toward leveraging foundation models for literature mining and observatory automation in astronomy \citep[e.g.,][]{2026ApJS..283...30S}. 
 
In this work, we advance the \texttt{RADAR} framework in three directions. 
First, we revisit the LLM at the heart of the GCN parser, evaluating three language models against the results of \citet{2025ApJS..280...71P} to quantify how the fidelity of automated metadata extraction improves as the underlying models mature (\autoref{sec:llm}). Second, we address the main architectural bottleneck in \texttt{RADAR}'s federated light-curve modeling (\autoref{sec:radioafterglow}). The original single-threaded likelihood dispatch over the \texttt{Octopus} messaging fabric was $\sim 40\times$ slower than local parallel sampling; by adding server-side thread pools and site-side subprocess pools for concurrent likelihood evaluation, we restore throughput comparable to centralized parallel \texttt{emcee} runs. We also benchmark the \textsc{JAX}-based surrogate \texttt{FIESTA} \citep{fiestaem} as a possible drop-in replacement for \texttt{afterglowpy}. Third, we extend \texttt{RADAR} beyond interpretation and inference into the planning of observations themselves, introducing an LLM-driven scheduling agent that translates natural-language observing requests into submission-ready scheduling blocks for the Karl~G.\ Jansky Very Large Array (VLA), which we validate through a scoring scheme and by verifying successful uploads to the Observation Preparation Tool  (OPT; \autoref{sec:VLASB}) of the National Radio Astronomy Observatory (NRAO). Taken together, we show how these developments push \texttt{RADAR} toward a faster and more fully automated path from GW alert to schedulable radio follow-up observations (\autoref{sec:conclusion}).

\newcommand{\mcell}[2]{{$#1$}{$\,\pm\,$}{$#2$}}
\begin{table*}
\begin{center}
\caption{Performance metrics of LLM-powered GCN parsers. Results for GPT-4.1 and
Llama-3.1-405B-Instruct are from \citet{2025ApJS..280...71P}, and other results are obtained in this paper (see \autoref{sec:llm} for details).
Mean values and standard deviations are computed over ten independent runs. The GCN-level recall $R_{\mathrm{GCN}}$ is denoted GCN-$R$ in \citet{2025ApJS..280...71P}. {Bold numbers are the highest performance values among all models for each performance metric.}}
\label{tab:metrics}
\begin{tabular}{lcccc}
\hline\hline
Model & $P$ $\uparrow$ & $R$ $\uparrow$ & $F_1$ $\uparrow$ & $R_{\mathrm{GCN}}$ $\uparrow$ \\
\hline
\texttt{GPT-4.1}        & \mcell{0.819}{0.023} & \mcell{0.727}{0.047} & \mcell{0.770}{0.035} & \mcell{0.724}{0.033} \\
\texttt{Llama-3.1-405B-Instruct} & \mcell{0.512}{0.022} & \mcell{0.447}{0.019} & \mcell{0.477}{0.021} & \mcell{0.375}{0.018} \\
\hline
\texttt{GPT-5.5}          & \mcell{0.935}{0.008} & \mcell{\mathbf{0.854}}{0.014} & \mcell{\mathbf{0.893}}{0.010} & \mcell{\mathbf{0.794}}{0.013} \\
\texttt{Claude-Opus-4.7}  & \mcell{\mathbf{0.978}}{0.014} & \mcell{0.796}{0.019} & \mcell{0.878}{0.016} & \mcell{0.721}{0.015} \\
\texttt{Gemini-3.5-Flash} & \mcell{0.930}{0.015} & \mcell{0.817}{0.013} & \mcell{0.870}{0.011} & \mcell{0.768}{0.021} \\
\hline
\end{tabular}
\end{center}
\end{table*}

\section{LLM-powered GCN parser} 
\label{sec:llm}

We evaluate the ability of LLMs to reproduce the structured radio-observation catalog of the GW170817 EM counterpart compiled from GCN Circulars, using the human-curated catalog of \citet{2025ApJS..280...71P} (their Table 3) as ground truth. We evaluate three LLMs: \texttt{GPT-5.5} (OpenAI), \texttt{Claude-Opus-4.7} (Anthropic), and \texttt{Gemini-3.5-Flash} (Google). For each provider, we selected the most recent model to which we had access at the time of this study. We note that \texttt{GPT-5.5} and \texttt{Claude-Opus-4.7} are their providers' flagship models, whereas \texttt{Gemini-3.5-Flash} belongs to Google's faster, lower-cost tier; the corresponding flagship, \texttt{Gemini-3.5-Pro}, was not available to us. 
All models were restricted to retrieve information from GCN Circulars alone {by querying their respective APIs with no tool use enabled (no web search, browsing, or external retrieval). Detailed model snapshots are given in \autoref{app:llm-config}.}
The performance of each LLM was measured following a procedure identical to that adopted by \citet{2025ApJS..280...71P}, which we briefly summarize below. {Following that work, an event here denotes a single radio observation reported in a Circular, specified by its frequency, flux density or upper limit, detection flag, and position; it does not denote an astrophysical transient. Every Circular in the corpus reports a follow-up of the same source, the optical counterpart of GW170817, so a Circular containing several events is one reporting several measurements of that source, for instance at different frequencies or epochs.} We define two levels of matching. Event-level matching requires that each individual observation parsed from a GCN Circular exactly reproduce the ground truth derived from expert-assigned values. GCN-level matching extends this criterion to the full set of events reported in a single Circular, such that a GCN Circular is considered a match only if all of its observations have corresponding matched LLM parsed observations.

To measure these matches, we define the following metrics with
$N_{\mathrm{AI}}$ denoting the number of events returned by the AI parser,
$N_{\mathrm{GT}}$ the number of ground-truth events, and $N_{\mathrm{match}}$
the number of parser events satisfying the event-level criterion. Event match
precision $P$, recall $R$, and $F_1$ score are then
\begin{equation}
    P = \frac{N_{\mathrm{match}}}{N_{\mathrm{AI}}}, \qquad
    R = \frac{N_{\mathrm{match}}}{N_{\mathrm{GT}}}, \qquad
    F_1 = \frac{2PR}{P + R}.
\end{equation}

Using $N^{\mathrm{GCN}}_{\mathrm{match}}$ to denote the number of Circulars
satisfying the GCN-level criterion and $N^{\mathrm{GCN}}_{\mathrm{tot}}$ for the total number
of Circulars considered, GCN match recall is
\begin{equation}
    R_{\mathrm{GCN}} = \frac{N^{\mathrm{GCN}}_{\mathrm{match}}}{N^{\mathrm{GCN}}_{\mathrm{tot}}}.
\end{equation}

We benchmark the three LLMs against the original \texttt{RADAR} prompt set \citep{2025ApJS..280...71P}, which we replicate in~\autoref{app:prompts}. Our results, reported in \autoref{tab:metrics}, show that the best-performing model, \texttt{GPT-5.5}, attains an event-match $F_1$ score of $0.893\pm0.010$, a 16\% improvement over the previous \texttt{GPT-4.1} results presented in \cite{2025ApJS..280...71P}, with substantially reduced variance. It also leads on the stricter GCN-level criterion, reaching $R_{\mathrm{GCN}} = 0.794\pm0.013$, a 10\% improvement over \texttt{GPT-4.1} and ahead of \texttt{Gemini-3.5-Flash} ($0.768\pm0.021$) and \texttt{Claude-Opus-4.7} ($0.721\pm0.015$). This metric is essential because the Circular, not the individual observation, is the unit \texttt{RADAR} ingests: a single mis-parsed observation obliges a human to review the whole Circular, so $R_{\mathrm{GCN}}$ measures how much of the incoming stream can be absorbed without intervention.

The LLM prediction mismatches are concentrated in a single failure mode. Of the 96 ground-truth events, 81 belong to GCN Circulars reporting more than one event, and these account for essentially all of the loss: the 15 single-event GCN Circulars are recovered in full by \texttt{Claude-Opus-4.7} and \texttt{Gemini-3.5-Flash} and in $97.3\pm3.4\%$ of cases by \texttt{GPT-5.5}, whereas recovery within multi-event Circulars falls to $88.9\pm1.2\%$ for \texttt{GPT-5.5}, $85.2\pm1.5\%$ for \texttt{Gemini-3.5-Flash}, and $77.8\pm1.8\%$ for \texttt{Claude-Opus-4.7}. The models under-enumerate rather than mis-read, for example, a GCN Circular reporting seven events may yield two. Field-level extraction is otherwise accurate: where a reported event corresponds to a ground-truth entry, the numeric differences are small; no unit-conversion error happens; nor is any source name, right ascension, or declination mismatched in any run. 

It should be noted that one caveat applies to any benchmark built on GW170817. As its GCN Circulars have been public since 2017, and \citet{2025ApJS..280...71P}, whose Table 3 is our ground truth, is itself published, whether that paper entered the training corpora of the newly benchmarked models is not disclosed and is unknown to us. With retrieval and other external tools disabled, whatever these models know of this event resides in their weights, and we cannot exclude that the answer key is among it. Two observations nonetheless argue against memorization dominating the newer models' performance. First, the prompts frame an extraction task on supplied text, asking the model to report only what the given GCN Circular states, to disregard other Circulars, and to omit fields the text does not provide (\autoref{app:prompts}); returning the exact frequency, flux density, detection flag, and position for that one Circular in a strict JSON schema is a different operation from reciting an aggregate table. Second, the scores are far from saturated, with the best model still leaving roughly one GCN Circular in five imperfect. GCN Circulars from future BNS mergers or other multi-messenger transients with extensive follow-up observations may offer opportunities for decisive tests, especially given the upcoming LIGO-Virgo-KAGRA Intermediate Run 1 (IR1\footnote{\url{https://observing.docs.ligo.org/plan/}}) and the recent start of operations of the Rubin Observatory Legacy Survey of Space and Time \citep[LSST;][]{2019ApJ...873..111I}.

A complementary application of LLMs to the same corpus is that of \citet{2026ApJS..283...30S}, who combine neural topic modeling with a retrieval-augmented \texttt{Mistral-7B-Instruct} pipeline to classify the full archive of more than $40{,}500$ GCN Circulars by messenger and observing band, and to extract GRB redshifts with $97.2\%$ accuracy on redshift-bearing circulars. Their accuracies are not directly comparable to the scores in \autoref{tab:metrics}, because their extraction targets a single scalar quantity per circular, whereas an event-level match here requires that frequency, flux density or upper limit, detection flag, and sky position simultaneously reproduce the human-curated value. Therefore, the two efforts are complementary: theirs demonstrates archive-scale triage and topic discovery with open-weight models, and ours shows the field-level fidelity needed to assemble a science-ready radio light curve for afterglow inference.

\begin{figure*} \centering \includegraphics[width=\textwidth]{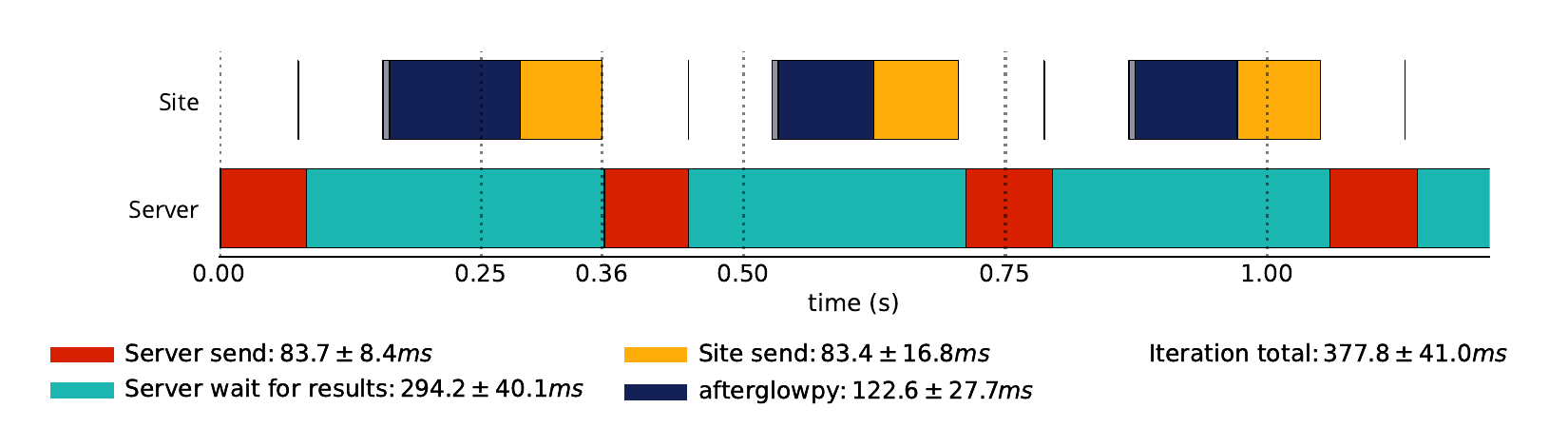} \caption{Profiling trace of a single \texttt{RADAR} MCMC likelihood evaluation cycle. Times are relative to an arbitrary reference point during the chain. The labelled durations report the mean $\pm$ standard deviation over the full MCMC run. Only operations contributing appreciably to the cycle time are shown. The communication overhead (server send, server wait, site send) is approximately twice the \texttt{afterglowpy} model evaluation time.} 
\label{fig:perf-trace} 
\end{figure*} 

\section{Accelerated Inference for Radio Modeling} 
\label{sec:radioafterglow} 

In this section we identify and address an important architectural bottleneck in \texttt{RADAR}.
A key evaluation metric for AI-driven multi-messenger frameworks is the latency with which data from heterogeneous observatories are acquired, processed, and collated to yield scientifically valuable constraints on model parameters. In the GW170817 case study, \texttt{RADAR} achieved an end-to-end latency of a few hours. However, this aggregate figure masks the disparate contributions of individual components. The GW inference, GCN parsing, and inter-module communication via \texttt{Octopus} each completed within minutes, whereas the federated radio light-curve fit, a distributed log-likelihood Markov
Chain Monte Carlo (MCMC) analysis using \texttt{afterglowpy}~\citep{2020ApJ...896..166R} with a Gaussian structured-jet model, required approximately 3--4~hr and dominated the total latency. Hereafter, we describe how we have identified the origin of this bottleneck and the strategy we have developed to mitigate it.

\subsection{Diagnosing the distributed MCMC bottleneck}
\label{sec:bottleneck} 

On a 16-core machine, the original \texttt{RADAR} code achieved an MCMC sampling rate of ${\approx}\,3.2$~samples\,s$^{-1}$ for the GW170817 radio data reported in \citet{2025ApJS..280...71P}, whereas a local \texttt{emcee}~\citep{foreman2013emcee} run with identical model and parameter settings sustained ${\approx}\,123$~samples\,s$^{-1}$, a factor of ${\approx}\,40\times$ difference. An analysis requiring ${\sim}\,4$~hr in \texttt{RADAR} would thus complete in ${\sim}\,6$~min locally. We attribute this penalty to two factors: (i)~the original \texttt{RADAR} implementation dispatched likelihood evaluations in a single thread, effectively serializing the computation; and (ii)~the \texttt{Octopus}/Apache Kafka publish-subscribe fabric introduces network round-trip latency on every likelihood call. 

Regarding (i) above, it is important to realize that the  \texttt{emcee} sampler evolves an ensemble of quasi-independent Markov chains (``walkers'') whose proposals depend on the collective walker positions but whose likelihood evaluations are mutually independent~\citep{2010CAMCS...5...65G}. This structure allows for trivial parallelization of the walker likelihoods in an MCMC iteration. In Python, however, the Global Interpreter Lock (GIL) serializes bytecode execution across threads within a single interpreter process, so that thread-based parallelism yields limited speedup for CPU-bound workloads. Subprocess-based parallelism, which spawns independent interpreter instances, circumvents the GIL and delivers near-linear scaling for embarrassingly parallel tasks such as per-walker likelihood evaluation. 

Regarding (ii) above, it is essential to note that the \texttt{RADAR} distributed MCMC architecture comprises a \emph{server} process that manages the sampler state, distributes likelihood requests, and aggregates results, together with one or more \emph{site} processes, running on separate machines, that evaluate likelihoods against their local data. Communication proceeds through the \texttt{Octopus} fabric: the server publishes model parameters, each site computes and publishes a partial log-likelihood, and the server collects and sums the results. Each evaluation cycle therefore involves four message transits through the fabric (two per direction). \autoref{fig:perf-trace} shows a profiling trace of a few such cycles. The combined communication overhead (${\approx}\,250$~ms) is roughly twice the \texttt{afterglowpy} model evaluation time (${\approx}\,120$~ms), confirming that network latency, not model cost, is the dominant bottleneck. Because this overhead arises from network propagation delays rather than computational work, it can be mitigated through concurrency. Further latency reduction by employing point-to-point communication between server and sites is possible, but would require bespoke network services capable of traversing institutional firewalls, adding substantial engineering complexity. We therefore retain the publish-subscribe architecture and instead overlap communication with computation, as described in detail in the next section. 

\begin{figure*} \centering \includegraphics[width=0.58\textwidth]{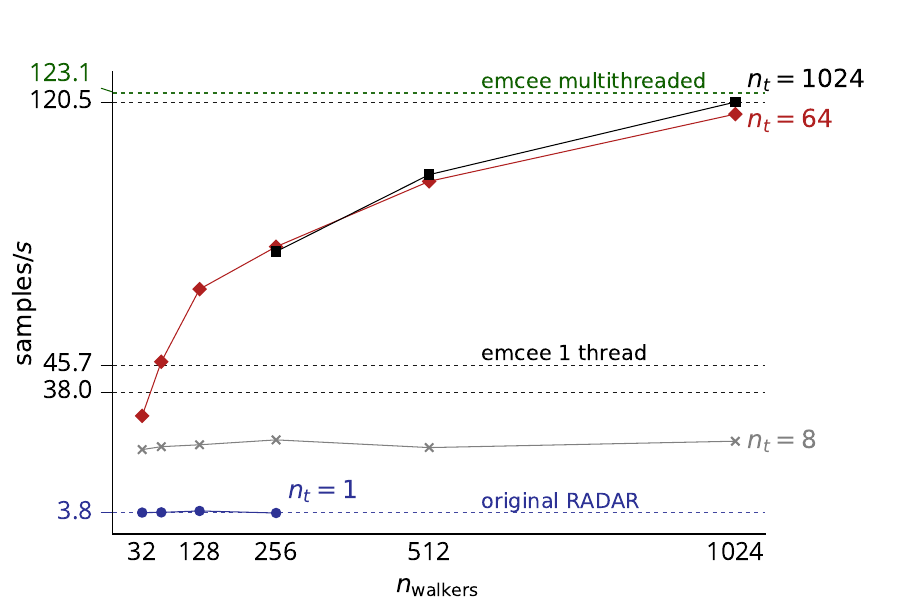} \caption{Sampling rate of the updated \texttt{RADAR} code as a function of the number of ensemble walkers ($n_{\rm walkers}$) and server-side threads ($n_{\rm t}$) on a 16-core machine. Horizontal lines indicate reference configurations: from bottom to top, original \texttt{RADAR} performance, $10\times$ original \texttt{RADAR} performance, single-thread \texttt{emcee}, updated \texttt{RADAR} performance for $n_{\text{walkers}} = n_t = 1024$, and multiprocess \texttt{emcee}. The light-curve model is \texttt{afterglowpy} with a Gaussian jet. For $n_{\rm walkers} = 1024$ and $n_{\rm t} \geq 64$, the distributed \texttt{RADAR} throughput matches the local parallel baseline.}
\label{fig:scaling} 
\end{figure*} 

\subsection{Concurrent likelihood evaluation} 
\label{sec:concurrent}

In order to parallelize the MCMC distributed loop, we introduce a number of modifications to the original \texttt{RADAR} architecture. On the server side, we replace the single-threaded remote likelihood evaluation stub with a multi-threaded version using a thread pool of size $n_{\rm t}$. The \texttt{emcee} sampler that drives the MCMC loop then issues up to $\min(n_{\rm walkers},\, n_{\rm t})$ likelihood requests in parallel. These requests are forwarded concurrently to the sites using the \texttt{Octopus} event fabric. The server threads then independently await the responses from the sites. This respectively overlaps both the \textbf{server send} and \textbf{server wait} phases of~\autoref{fig:perf-trace} across walkers. A reduction step then combines the individual site likelihoods into a global likelihood. On the site side, requests received through \texttt{Octopus} are forwarded immediately to a subprocess pool, which allows parallel evaluation of the \texttt{afterglowpy} model that is free of GIL contention.

\autoref{fig:scaling} shows the resulting sampling rate as a function of $n_{\rm walkers}$ and $n_{\rm t}$ on a 16-core machine. For $n_{\rm walkers} = 1024$ and $n_{\rm t} \geq 64$, the \texttt{RADAR} throughput reaches ${\approx}\,120$~samples\,s$^{-1}$, comparable to a local parallel \texttt{emcee} run on the same hardware (${\approx}\,123$~samples\,s$^{-1}$). The ${\approx}\,40\times$ penalty of the original implementation is thereby eliminated without sacrificing data locality or the privacy-preserving properties of the federated architecture. A complementary scaling analysis on the site side confirms that the optimal subprocess pool size matches the number of physical CPU cores, as expected for a CPU-bound workload that cannot benefit from parallelism beyond the available hardware concurrency.

We note that increasing the number of walkers also influences the burn-in 
cost of the sampler. In a collection of fully independent chains, each 
walker is initialized at a random point in parameter space and typically 
requires an initial transient phase before reaching the region of 
appreciable posterior support. Samples generated during this initial 
phase are not representative of the target distribution and are 
therefore discarded as burn-in. Under the assumption that walkers 
evolve independently, the total number of discarded samples scales 
linearly with $n_{\rm walkers}$, so increasing the walker count 
increases the aggregate burn-in cost. For ensemble samplers, however, 
this scaling need not hold in practice. The inter-walker coupling 
that enables the affine-invariant proposal can reduce the burn-in 
required per walker as $n_{\rm walkers}$ increases~\citep{doi.org/10.1111/anzs.12358}, thereby partially 
compensating for the larger ensemble size. Determining the optimal 
balance between walker count and burn-in length for the \texttt{RADAR} 
parameter space will require a dedicated convergence analysis and is 
beyond the scope of the present work. Accordingly, the throughput 
values reported above include all generated samples.

\begin{figure*} 
\centering \includegraphics[width=0.92\textwidth]{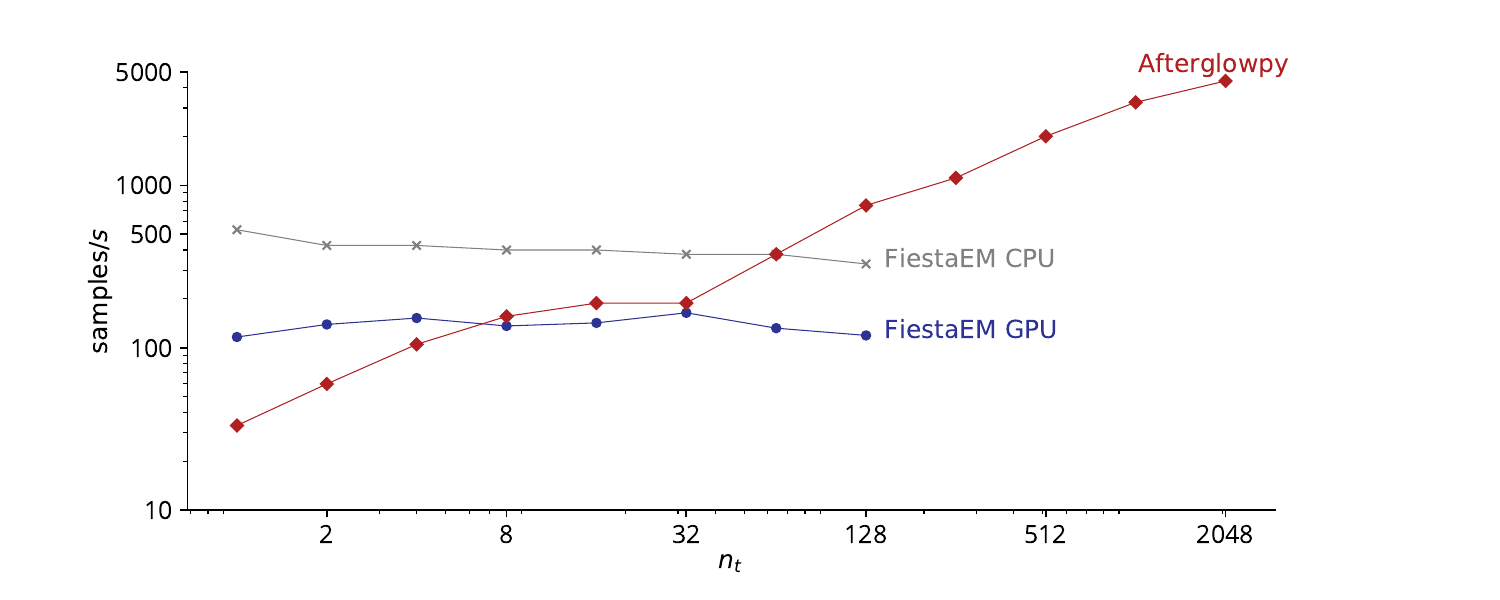} 
\caption{Sampling rate as a function of pool size ($n_{\rm t}$) for \texttt{FIESTA} (GPU and CPU backends) and \texttt{afterglowpy}. \texttt{FIESTA} uses thread pools; \texttt{afterglowpy} uses subprocess pools. The number of ensemble walkers is $\max(n_{\rm t},\, 32)$. \texttt{FIESTA} outperforms \texttt{afterglowpy} at $n_{\rm t} = 1$ but cannot match its near-linear subprocess scaling due to GIL contention under thread-based parallelism. The GPU backend underperforms the CPU backend owing to kernel-launch and transfer overhead at the small per-call batch sizes used here. Benchmarks were performed on the \texttt{DeltaAI} supercomputer (NVIDIA H100 GPUs); consistent results were obtained on the \texttt{Polaris} supercomputer (NVIDIA A100 GPUs).} 
\label{fig:fiesta} 
\end{figure*} 

\subsection{Benchmarking \texttt{FIESTA}: A surrogate light-curve model} 
\label{sec:fiesta} 

As a complementary approach to reducing the per-evaluation cost of the radio light-curve modeling, we benchmarked \texttt{FIESTA}~\citep{fiestaem}, an open-source \texttt{Python} library that trains neural-network surrogates to emulate established afterglow codes. We chose \texttt{FIESTA} for 
three reasons. 
First, it is fully open source and publicly available, which ensures 
reproducibility and allows straightforward integration into the 
\texttt{RADAR} codebase without licensing constraints. Second, \texttt{FIESTA} benefits from an active and responsive development team, which enabled us to resolve the compatibility and configuration issues encountered during benchmarking on short timescales. Third, 
and most importantly, \texttt{FIESTA} employs neural network 
surrogates trained to emulate semi-analytic afterglow codes such 
as \texttt{afterglowpy}. 
The profiling analysis of~\autoref{sec:bottleneck} identified likelihood 
evaluation cost as the second significant contributor to overall latency. Once the communication overhead was addressed, model evaluation cost became the next limiting factor, hence
a surrogate model that reduces this cost was a natural candidate for
further improving the performance of \texttt{RADAR}'s federated 
MCMC loop.

\texttt{FIESTA} is built on \texttt{JAX}~\citep{jax}, which provides a high-performance, GPU-capable \texttt{NumPy}-compatible API with just-in-time (JIT) compilation. We used the \texttt{afgpy\_gaussian\_CVAE} surrogate, which emulates the Gaussian structured-jet model of \texttt{afterglowpy} (\texttt{jetType\,=\,jet.Gaussian}). Benchmarks were performed on the \texttt{DeltaAI} supercomputer~\citep{delta-ai} at NCSA/UIUC (4 NVIDIA H100 GPUs, 288 ARM cores per node) using exclusively allocated compute nodes and the Cray-provided Python~3.11 environment. We ran \texttt{emcee} with pool sizes ranging from 1 to 2048 threads or subprocesses. Because \texttt{JAX} is incompatible with Python's \texttt{multiprocessing} module (forking a process that holds an active CUDA context leads to undefined behavior), \texttt{FIESTA} benchmarks used thread pools, whereas \texttt{afterglowpy} benchmarks used subprocess pools. For the CPU-only \texttt{FIESTA} runs, we disabled GPU access by removing the CUDA Python packages, forcing \texttt{JAX} to select its CPU backend. In all cases, the number of ensemble walkers was set to $\max(n_{\rm t},\, 32)$, where 32 is the minimum permitted by \texttt{emcee} for the dimensionality of the model. \autoref{fig:fiesta} presents the results. Three findings are noteworthy:
\begin{itemize}
\item {\it Single-threaded performance}--With a single thread, \texttt{FIESTA} (CPU) evaluates the surrogate model at ${\approx}\,5\times$ the rate of \texttt{afterglowpy} (both after JIT warm-up), confirming the expected advantage of a neural-network surrogate over a semi-analytic integration. 
\item {\it Multiprocess scaling}--\texttt{afterglowpy} exhibits near-linear speedup with the number of subprocesses, saturating only when the physical core count is reached. \texttt{FIESTA}, confined to thread-level parallelism by the \texttt{JAX}/multiprocessing incompatibility, shows \emph{decreasing} throughput as the thread count is increased. This degradation is consistent with GIL contention: although \texttt{JAX} releases the GIL during compiled kernels, the Python-level orchestration code (array construction, callback dispatch, result collection) reacquires it, serializing an increasing fraction of the work as more threads compete. 
\item {\it GPU versus CPU}--The GPU-accelerated \texttt{FIESTA} backend is slower than the CPU backend across all thread counts tested. We attribute this to the combination of small per-call batch sizes (a single light curve per likelihood evaluation), which fail to amortize GPU kernel launch and host-device transfer overhead, and the additional GIL contention introduced by the CUDA runtime. Batching multiple walker evaluations into a single vectorized \texttt{JAX} call, using \texttt{jax.vmap} rather than a Python-level thread pool, would be the natural remedy; we leave this to future work. 
\end{itemize}

We obtained similar results on the \texttt{Polaris} supercomputer at Argonne National Laboratory (4 NVIDIA A100 GPUs, 32 AMD Zen~3 cores per node), confirming that the scaling behavior is not specific to a single hardware platform. 

In summary, \texttt{FIESTA} offers a significant single-threaded evaluation speedup over \texttt{afterglowpy}, but realizing this advantage with parallel samplers will require \texttt{JAX}-native vectorization strategies (e.g., \texttt{jax.vmap} over walkers) that bypass the Python threading model entirely. Until such an integration is achieved, \texttt{afterglowpy} with subprocess-parallel \texttt{emcee} remains the more performant choice for production \texttt{RADAR} runs.

\section{LLM-driven VLA scheduling} 
\label{sec:VLASB}

The VLA is a PI-driven U.S.\ national radio facility operated by the NRAO \citep[][]{Perley2011}. As the most sensitive, highest-resolution radio interferometer currently operating in the U.S., with broad frequency coverage, it is a powerful instrument for studying astrophysical transients and their radio counterparts \citep[e.g.,][]{2020SSRv..216...81A,2001Natur.410..338B,2005Natur.434.1112C,2012ApJ...746..156C,2016ApJ...830...42C,2023ApJ...953..179C,2016ApJ...832L...1D,2012ApJ...747...70F,Hallinan_2017,1998Natur.395..663K,2025PASP..137h4102S,2006Natur.442.1014S,2026ApJ..1002..194O}. The VLA has made major contributions to multi-messenger astronomy (MMA), from the discovery of the radio afterglow of GW170817 \citep{2017Sci...358.1579H} to the recent identification of the radio counterpart to SN\,2025ulz \citep{2026arXiv260405128O}, a transient temporally and spatially compatible with the sub-threshold BNS candidate S250818k \citep{2025GCN.41437....1L,2025ApJ...995L..27G,2025ApJ...995L..59K,2026arXiv260502639A,2026ApJ..1001L..20H,2026ApJ...996L..24Y}.

In this Section, we describe how we develop a LLM-driven framework to convert natural-language observing requests into submission-ready scheduling blocks (SBs) for the VLA. We also develop a framework for scoring LLM-generated SBs against human-generated ones to check their compliance with basic observing strategies and requirements. 

\subsection{\texttt{SAGE}: Scheduling-block Automated Generation Engine}
\label{sec:sage}

VLA observations are organized into SBs\footnote{\url{https://science.nrao.edu/facilities/vla/docs/manuals/opt-manual/text-files-and-catalogs-opt/importing-sbs}}, the fundamental units in which time is requested, validated, and executed. Once submitted through the NRAO's OPT, each SB is validated in two stages: an automated consistency check at submission, followed by a manual NRAO review for operational errors not caught automatically (e.g., an inappropriate antenna wrap, missing scans, or misassigned intents). Approved SBs are converted into machine-readable observing scripts and entered into the dynamic VLA scheduling queue, where execution is governed by observing priority and weather constraints, principally wind speed and the atmospheric phase limit. Because SB creation is both a barrier for users new to radio astronomy and a time-consuming step even for experienced VLA observers, here we focus on automating SB production for the general user, enabling natural-language commands to generate SBs that can be uploaded directly into the OPT.

A VLA SB can be created and uploaded to the OPT independently if it follows a valid format, which includes the declarations the OPT requires to build a new SB: the version, source and hardware catalogs, SB header, and the execution block containing the targets and observing loops. LLM-powered SB creation can therefore be posed as a template-completion problem, in which the LLM must generate an OPT file while preserving the observing logic it is trained on. We define this template in \autoref{tab:sb_ground_truths}, listing each major SB component, its purpose, and the information it must contain to pass validation.

We began by using \texttt{GPT-5.5} in ChatGPT with the reasoning-effort setting set to high, and web access enabled so as to provide the LLM with the \href{https://science.nrao.edu/facilities/vla/docs/manuals/oss/referencemanual-all-pages}{VLA Observational Status Summary 2025B Complete Manual}, which summarizes the components and risks involved in creating a valid SB. We also provided specified sections within the VLA manual and relevant VLA pages that contain useful information the LLM should consider when generating an SB. For a more detailed description of the reference material, see \autoref{lst:training}. 

Before a complete VLA SB can be generated, the user must create an SCT.txt file (see \autoref{lst:SRC_CAT_skeleton}) listing the names and positional coordinates of the targets, and import it into the Source Catalog Tool (SCT) within the OPT. The catalog name is set by the user (see \autoref{lst:LLM-SB prompt}) and must also appear in the SRC-CAT declaration line of the SB (see \autoref{lst:sb_skeleton}). Hence, we first created and uploaded a catalog named ``test1''. Then, we selected two representative SBs that had been created manually for a single X-band observation of SN\,2024rjw and a multi-band observation of SN\,2021bmf under VLA program VLA/26A-349 (PI: O'Dwyer). These observations were carried out with the VLA in its A-array configuration at 3\,GHz (S-band receiver), 10\,GHz (X-band receiver), and 15\,GHz (Ku-band receiver). With these two template SBs in hand, we developed a set of prompts that ultimately enabled the LLM to reproduce the template SBs for SN\,2024rjw and SN\,2021bmf (marked as  dev-target in \autoref{tab:sb_performance} and \autoref{tab:sb_times_rms}). In this process, we guided the LLM incrementally: we first asked it to build a minimal calibration-only SB containing the flux calibrator (i.e., 3C286) but no target-observation loop, and then to produce a complete SB for a continuum observation (including the phase-calibrator slew and a repeating phase-calibrator–target loop). The resulting SB uploaded successfully to the OPT, demonstrating that the LLM, with some direction, could produce a valid SB for a VLA continuum observation. A shorthand version of the SB.txt file is shown in~\autoref{lst:sb_skeleton}, highlighting the main components the LLM-generated SB is expected to have. 

Next, we created a corresponding ``Scheduling-block Automated Generation Engine'' (\texttt{SAGE}). \texttt{SAGE} consists of the frozen prompt for \texttt{GPT-5.5} High reproduced in Appendix~\autoref{lst:LLM-SB prompt}, invoked manually through the ChatGPT interface for each SB. We then tested the performance of \texttt{SAGE} for SB production   (see \autoref{sec:SBresults}). We note that \texttt{SAGE} could be further developed into an API-based application or an autonomous software agent.

\begin{table*}
\centering
\caption{Criteria used to score LLM-generated VLA SBs. 
The 17 ground-truth categories map to the 21 individually scored criteria of \autoref{tab:scoring}. See text for discussion (\autoref{sec:score}).
}
\label{tab:sb_ground_truths}

\scriptsize
\setlength{\tabcolsep}{4pt}
\renewcommand{\arraystretch}{1.45}

\begin{tabular}{c|c|c}
    \hline
    \textbf{Ground-truth category} &
    \textbf{Scoring category (\autoref{tab:scoring})} &
    \textbf{Pass criterion} \\
    \hline

    \parbox[c]{1.5in}{%
        \strut\raggedright
        \textbf{Version declaration}
        \strut
    } &
    \parbox[c]{2.05in}{%
        \strut\raggedright
        \texttt{version\_present}
        \strut
    } &
    \parbox[c]{2.65in}{%
        \strut\raggedright
        \texttt{VERSION; 7;} must appear as the first non-blank line of the generated SB.
        \strut
    } \\[4pt]
    \hline

    \parbox[c]{1.5in}{%
        \strut\raggedright
        \textbf{Source catalog declaration}
        \strut
    } &
    \parbox[c]{2.05in}{%
        \strut\raggedright
        \texttt{SRC\_CAT\_present}
        \strut
    } &
    \parbox[c]{2.65in}{%
        \strut\raggedright
        A valid \texttt{SRC-CAT; (personal catalog), VLA;} must appear in the generated SB, VLA should always appear.
        \strut
    } \\[4pt]
    \hline

    \parbox[c]{1.5in}{%
        \strut\raggedright
        \textbf{Hardware catalog declaration}
        \strut
    } &
    \parbox[c]{2.05in}{%
        \strut\raggedright
        \newline\texttt{HDWR\_CAT\_present}
        \newline\texttt{hardware\_in\_allowlist}
        \strut
    } &
    \parbox[c]{2.65in}{%
        \strut\raggedright
        A valid \texttt{HDWR-CAT; NRAO Defaults;} line must be present in the SB header below the source catalog declaration. Each SB scan should have a hardware receiver declared, and the name must exactly match a valid VLA OPT receiver and ensure that receiver choice matches the telescope configuration.
        \strut
    } \\[4pt]
    \hline

    \parbox[c]{1.5in}{%
        \strut\raggedright
        \textbf{Loop closure}
        \strut
    } &
    \parbox[c]{2.05in}{%
        \strut\raggedright
        \texttt{loop\_closure}
        \strut
    } &
    \parbox[c]{2.65in}{%
        \strut\raggedright
        Every \texttt{LOOP-START} must have a corresponding
        \texttt{LOOP-END}.
        \strut
    } \\[4pt]
    \hline

    \parbox[c]{1.5in}{%
        \strut\raggedright
        \textbf{Initial setup scan}
        \strut
    } &
    \parbox[c]{2.05in}{%
        \strut\raggedright
        \texttt{setup\_scans\_first}
        \strut
    } &
    \parbox[c]{2.65in}{%
        \strut\raggedright
        Initial setup scan should occur before target scans. For 3-bit, 8-bit, or tuning changes, setup scans must be included as required by the hardware configuration. \texttt{Dummy, Slew, Attenuation, and Requantizer gain} scans.
        \strut
    } \\[4pt]
    \hline

    \parbox[c]{1.5in}{%
        \strut\raggedright
        \textbf{Scan intents}
        \strut
    } &
    \parbox[c]{2.05in}{%
        \strut\raggedright
        \texttt{intents\_valid}
        \strut
    } &
    \parbox[c]{2.65in}{%
        \strut\raggedright
        Scan intents must match the role of each scan, including CalFlux, CalBP, CalGain, ObsTgt, and setup intents.
        \strut
    } \\[4pt]
    \hline

    \parbox[c]{1.5in}{%
        \strut\raggedright
        \textbf{Scan ordering}
        \strut
    } &
    \parbox[c]{2.05in}{%
        \strut\raggedright
        \texttt{template\_order}
        \strut
    } &
    \parbox[c]{2.65in}{%
        \strut\raggedright
        Scan order must follow the expected sequence:
        setup $\rightarrow$ flux/bandpass $\rightarrow$
        phase calibrator $\rightarrow$ target.
        \strut
    } \\[4pt]
    \hline

    \parbox[c]{1.5in}{%
        \strut\raggedright
        \textbf{Flux calibration}
        \strut
    } &
    \parbox[c]{2.05in}{%
        \strut\raggedright
        \texttt{flux\_present}
        \newline\texttt{flux\_cal\_standard}
        \strut
    } &
    \parbox[c]{2.65in}{%
        \strut\raggedright
        At least one scan must contain the \texttt{CalFlux} intent. The flux calibrator must be a standard flux calibrator with a valid CASA/AIPS model:
        3C286, 3C48, or 3C147. Possible flares should also be considered; see the \href{https://science.nrao.edu/facilities/vla/docs/manuals/oss/performance/fdscale}{VLA flux-density scale documentation}
        \strut
    } \\[4pt]
    \hline

    \parbox[c]{1.45in}{%
        \strut\raggedright
        \textbf{Bandpass calibration}
        \strut
    } &
    \parbox[c]{2.05in}{%
        \strut\raggedright
        \texttt{bandpass\_present}
        \strut
    } &
    \parbox[c]{2.65in}{%
        \strut\raggedright
        At least one scan must contain the \texttt{CalBP} intent.
        \strut
    } \\[4pt]
    \hline

    \parbox[c]{1.5in}{%
        \strut\raggedright
        \textbf{Phase calibrator}
        \strut
    } &
    \parbox[c]{2.05in}{%
        \strut\raggedright
        \texttt{phase\_cal\_present}
        \newline\texttt{phase\_cal\_valid}
        \newline\texttt{phase\_cal\_separation}
        \strut
    } &
    \parbox[c]{2.65in}{%
        \strut\raggedright
        At least one scan must contain the \texttt{CalGain} intent and exist in the VLA calibrator catalog, be compact and bright enough for gain solutions. Must be observable in the selected band and array configuration; Quality Codes P or S are preferred. The target--phase-calibrator must have an angular separation within the recommended limits (see \autoref{sec:score}).
        \strut} \\[4pt]
    \hline

    \parbox[c]{1.5in}{%
        \strut\raggedright
        \textbf{Target scan}
        \strut
    } &
    \parbox[c]{2.05in}{%
        \strut\raggedright
        \texttt{target\_present}
        \strut
    } &
    \parbox[c]{2.65in}{%
        \strut\raggedright
        At least one scan with the \texttt{ObsTgt} intent
        must be present. Target name should exist in a catalog declared in the SRC-CAT line.
        \strut
    } \\[4pt]
    \hline

    \parbox[c]{1.5in}{%
        \strut\raggedright
        \textbf{Phase scan per target loop}
        \strut
    } &
    \parbox[c]{2.05in}{%
        \strut\raggedright
        \texttt{one\_phase\_per\_target\_loop}
        \strut
    } &
    \parbox[c]{2.65in}{%
        \strut\raggedright
        Every target loop must contain exactly one
        phase-calibrator line.
        \strut
    } \\[4pt]
    \hline

    \parbox[c]{1.5in}{%
        \strut\raggedright
        \textbf{Array configuration}
        \strut
    } &
    \parbox[c]{2.05in}{%
        \strut\raggedright
        \texttt{config\_matches\_date}
        \strut
    } &
    \parbox[c]{2.65in}{%
        \strut\raggedright
        The selected VLA configuration must match the
        configuration valid for the observing-date window.
        \strut
    } \\[4pt]
    \hline

    \parbox[c]{1.5in}{%
        \strut\raggedright
        \textbf{High-frequency reference pointing}
        \strut
    } &
    \parbox[c]{2.05in}{%
        \strut\raggedright
        \texttt{highfreq\_refpointing}
        \strut
    } &
    \parbox[c]{2.65in}{%
        \strut\raggedright
        High-frequency Ku, K, Ka, and Q band SBs should include X-band reference pointing or be explicitly waived in the prompt.
        \strut
    } \\[4pt]
    \hline
    
    \parbox[c]{1.5in}{%
        \strut\raggedright
        \textbf{Coordinate prompt SCT.txt consistency}
        \strut
    } &
    \parbox[c]{2.05in}{%
        \strut\raggedright
        \texttt{coord\_SCT\_match}
        \strut
    } &
    \parbox[c]{2.65in}{%
        \strut\raggedright
        The SCT.txt target coordinates must agree with the declared RA and DEC of the LLM prompt. 
        \strut
    } \\[4pt]
    \hline

    \parbox[c]{1.5in}{%
        \strut\raggedright
        \textbf{Phase-target cycle time}
        \strut
    } &
    \parbox[c]{2.05in}{%
        \strut\raggedright
        \texttt{cycle\_time}
        \strut
    } &
    \parbox[c]{2.65in}{%
        \strut\raggedright
        The phase-to-target cycle time should fall within the NRAO guidance for the receiver band and array configuration.
        \strut
    } \\[4pt]
    \hline

    \parbox[c]{1.5in}{%
        \strut\raggedright
        \textbf{Generated SB-duration}
        \strut
    } &
    \parbox[c]{2.05in}{%
        \strut\raggedright
        \texttt{Total\_duration}
        \strut
    } &
    \parbox[c]{2.65in}{%
        \strut\raggedright
        The total SB duration must not exceed the requested total SB time in the LLM-SB construction prompt. On-source time is counted by uploading the SB to the OPT and using \href{https://obs.vla.nrao.edu/ect/}{exposure calculator} to estimate observation sensitivity.
        \strut
    } \\[4pt]
    \hline

\end{tabular}
\end{table*}

\begin{table*}
\centering
\caption{Performance of \texttt{SAGE} evaluated
with \texttt{SCORE} for the SBs of the sources described in  \autoref{sec:SBresults} The score is defined as
$S=N_{\rm pass}=21-N_{\rm fail}$. Criteria that are flagged for human review are counted toward $N_{\rm flag}$. SN\,2021bmf and SN\,2024rjw are marked as dev-targets as they were used for prompt development in \texttt{SAGE} (\autoref{sec:sage}).}
\label{tab:sb_performance}
\begin{tabular}{lcccccccc}
\hline
Target & Band & $N_{\rm pass}$ & $N_{\rm fail}$ & $N_{\rm flag}$ & Score & Failed/Flagged criterion \\
 & & & & & & \\
\hline
GW170817 & S & 21 & 0 & 0 & 21 & None (but see discussion in \autoref{sec:SBresults}) \\
\hline
SN2026gzf & C & 21 & 0 & 0 & 21 & None \\
\hline
SN2025ulz & S & 21 & 0 & 0 & 21 & None (but see discussion in \autoref{sec:SBresults}) \\
\hline
SN2025ulz & Ku & 20 & 1 & 0 & 20 & Total SB duration failed to match requested one \\
\hline
SN2024rjw (dev-target) & X & 21 & 0 & 0 & 21 & None \\
\hline
SN2021bmf (dev-target) & S/X/Ku & 20 & 1 & 0 & 20 & Total SB duration failed to match requested one \\
\hline
\end{tabular}
\end{table*}

\begin{table*}
\begin{center}
\centering
\caption{VLA OPT comparison of LLM-generated (GPT-5.5) and corresponding ground truth (GrTr) SBs. The LST start time for the LLM-generated and ground truth SBs of SN\,2024rjw differ due to a difference in Flux and Phase calibrator choices made by the LLM. 
Target on-source times were obtained from the OPT static reports. The noise root-mean-square (RMS) sensitivities were calculated manually using the VLA 
Exposure Calculator and are reported in $\mu$Jy beam$^{-1}$. For an in depth discussion of the difference in ``Total SB Duration'' for the LLM-generated and ground truth SBs, and related scoring reported in \autoref{tab:sb_performance},  see \autoref{sec:score}.
\label{tab:sb_times_rms}}
\begin{tabular}{cccccccc}
\toprule
Source & Band & SB & Total SB & Static Report  &  On-Source & Hardware Resource & RMS \\
 &  &  & Duration &  LST & time &  &  \\
\midrule

GW170817 & S & GPT-5.5 & 03:31:00 & 10:52:30 & 02:28:19.175 & \texttt{S16f2B} & 3.1  \\
GW170817 & S & GrTr & 03:30:00 & 10:52:30 & 02:41:45.677 & \texttt{S16f3B} & 2.9  \\

\midrule

SN2026gzf & C & GPT-5.5 & 01:00:00 & 09:55:00 & 00:33:17.461 & \texttt{C32f2Amixalt-blank} & 2.9 \\
SN2026gzf & C & GrTr & 01:00:00 & 09:55:00 & 00:33:17.461 & \texttt{C32f2Amixalt} & 2.9  \\

\midrule

SN2025ulz & S & GPT-5.5 & 03:01:00 & 17:47:30 & 01:55:09.054 & \texttt{S16f3B} & 3.4  \\
SN2025ulz & S & GrTr & 03:00:00 & 17:47:30 & 02:10:03.930 & \texttt{S16f3B} & 3.3  \\

\midrule

SN2025ulz & Ku & GPT-5.5 & 03:28:50 & 10:10:00 & 02:15:08.107 & \texttt{Ku48f3DCB} & 1.8 \\
SN2025ulz & Ku & GrTr & 03:25:00 & 10:10:00 & 02:42:09.776 & \texttt{Ku48f3DCB} & 1.7  \\

\midrule

SN2024rjw (dev-target)& X & GPT-5.5 & 02:00:00 & 22:07:30 & 01:21:51.129 & \texttt{X32f2A} & 2.1 \\
SN2024rjw (dev-target)& X & GrTr & 02:00:00 & 16:55:00 & 01:17:39.124 & \texttt{X32f2A} & 2.1  \\

\midrule

SN2021bmf (dev-target)& S/X/Ku & GPT-5.5 & 02:00:20 & 16:50:00 & 00:17:40.058 & \texttt{S16f2A} & 9.0/4.0/5.2 \\
SN2021bmf (dev-target)& S/X/Ku & GrTr & 02:00:00 & 16:50:00 & 00:19:42.616 & \texttt{S16f2A} & 8.4/4.2/5.3  \\

\bottomrule
\end{tabular}
\end{center}
\end{table*}

\subsection{\texttt{SCORE}: Scheduling-block Compliance and Observation-Readiness Evaluator}
\label{sec:score}
In order to score the LLM-created SBs and quantitatively compare them against the ground truth, i.e., SBs created by VLA observers in our team to successfully observe a variety of transients, we created a new \texttt{RADAR} module dubbed ``Scheduling-block Compliance and Observation-Readiness Evaluator'' (\texttt{SCORE}), that we use to score the LLM-powered SB generator's performance based on the criteria described in \autoref{tab:sb_ground_truths}, and via the scoring scheme summarized in \autoref{tab:scoring}. 

\texttt{SCORE} tests the LLM's ability to select appropriate calibrators, hardware resources, scan sequences, cycle times, and reference-pointing procedures against a manually created, valid SB. Scoring is performed by parsing the SB and evaluating 17 ground-truth categories (see \autoref{tab:sb_ground_truths}), which map onto the 21 individually scored criteria listed in \autoref{tab:scoring}.

Sixteen of the 21 criteria perform mechanical checks, such as file version, catalog declarations, loop closure, initial setup scan, valid hardware names, scan intents, required calibrators, target scans, scan order, array-configuration compatibility, and high-frequency reference pointing. These 16 criteria are marked as ``Mechanical'' in \autoref{tab:scoring} and are scored as +1 (pass) or 0 (fail). 

Two criteria are marked as ``Computed'' in \autoref{tab:scoring}. The \texttt{Total\_duration} check compares the total duration of the generated SB with the total SB time requested in the user's prompt. The \texttt{coord\_SCT\_match} check verifies that the target coordinates (right ascension, R.A., and declination, Dec.) listed in the LLM-generated source catalog (i.e., the \texttt{SCT.txt} file) match those given in the user's prompt. Both of these checks are scored as +1 (pass) or 0 (fail). 

Two further criteria, marked as ``External'' in \autoref{tab:scoring} (\texttt{phase\_cal\_valid} and \texttt{phase\_cal\_separation}), require external information to validate the phase calibrator. There are two general guidelines for choosing a phase calibrator: (i) it should lie as close as possible to the target (within 10\,deg at L and S bands, 7\,deg at C and X bands, and 3\,deg at Ku band and above); and (ii) it should be point-like at the observing frequency and array configuration, and sufficiently bright, i.e., it should carry a calibrator quality code of either Primary (P) or Strong (S). Because multiple phase calibrators can satisfy (i) and (ii) for any given target, validating the choice of phase calibrator in \texttt{SCORE} requires an approach that goes beyond simply checking whether it is identical to the one used in the ground-truth SB. Moreover, when suitable calibrators are scarce for a given source and frequency, one may need to select a calibrator at an angular separation somewhat larger than recommended, or the information needed to verify (ii) may not be available. Hence, if the phase calibrator in the LLM-generated SB is identical to that of the ground-truth SB, both criteria are scored as +1; otherwise, \texttt{SCORE} checks whether the LLM-chosen calibrator satisfies criteria (i) and (ii) above. Each criterion that is satisfied is scored as +1, while each criterion that is violated or cannot be verified is provisionally flagged as -1 (rather than 0) to indicate that the SB requires manual inspection before a score can be assigned. After manual inspection, a pass (+1) or fail (0) score is assigned by hand.

The last criterion (\texttt{cycle\_time}; see \autoref{tab:scoring}) is marked as ``Advisory''. It compares the phase calibrator target cycle time with NRAO tabulated cycle times\footnote{\url{https://science.nrao.edu/facilities/vla/docs/manuals/obsguide/calibration\#CalibrationCycle}} that assume average weather conditions. As such, the NRAO tabulated cycle times should be regarded as recommendations rather than hard limits, and departures from them may be acceptable. This criterion is therefore scored as +1 when satisfied and provisionally flagged as -1 for manual inspection when it is not, after which a pass (+1) or fail (0) score is assigned by hand. In all cases, the -1 flag is provisional and is always resolved to +1 or 0 by manual inspection, so that the final \texttt{SCORE} of an SB ranges from 0 to 21.

\subsection{Results}
\label{sec:SBresults}

Using \texttt{SAGE} and \texttt{SCORE}, we evaluate the LLM's ability to reproduce SBs that were previously created by members of our team to conduct observations of two classes of astrophysical transients: (i) VLA continuum observations of core-collapse SNe, and (ii) deep VLA follow-up observations of EM counterparts found during the follow up of GW alerts,  and specifically GW170817 \citep{Coulter_2017, Hallinan_2017,2018Natur.561..355M} and SN\,2025ulz. The last was found in the error region of the low-confidence GW alert S250818k  \citep{2025GCN.41414....1S,2025ApJ...995L..59K,2026arXiv260405128O, 2026ApJ...996L..24Y, 2026arXiv260502639A}. 
To ensure robustness of our results, we scored LLM-generated SBs across different VLA receiver bands and observing programs:
\begin{itemize}
    \item A C-band (6\,GHz) observation of SN\,2026gzf (class i above), under VLA program VLA/25B-282, \citep[PI: O'Dwyer;][]{chen2026decadalpreexplosionactivitycircumstellar,2026GCN.44239....1O,2026ApJ..1007L...9R, 2026ApJ..1006L..13O};
    \item {\bf{Dev-target}} Multi-band (S/X/Ku or 3--15\,GHz) observation of SN\,2021bmf (class i above) under VLA program VLA/26A-349 (PI:O'Dwyer; \citealt{Anand_2024,srinivasaragavan2024opticalradioanalysissystematically},O'Dwyer et al. 2026 in prep);
    \item {\bf{Dev-target}} X-band (10\,GHz) observation of SN\,2024rjw (class i above) under VLA program VLA/26A-349 (PI:O'Dwyer; \citealt{2026ApJ..1002..194O}, O'Dwyer et al. 2026 in prep);
    \item S- (3\,GHz) and Ku-band (15\,GHz) observations of SN\,2025ulz (class ii above), under VLA program VLA/22B-235  \citep[PI: Corsi;][]{2025GCN.41414....1S,2025ApJ...995L..59K,2026arXiv260405128O, 2026ApJ...996L..24Y, 2026arXiv260502639A}.
    \item An S-band observation of GW170817 (class ii above), under VLA program VLA/24A-119 (PI: O'Dwyer; \citealt{Hallinan_2017}).
\end{itemize} 
We note that in this study we treat high- and low-frequency SBs separately and restrict the analysis to continuum observations, setting aside spectral-line, very-low-frequency, and polarization data. 

The LLM-generated SBs matched the structure and observing requirements of the corresponding ground-truth SBs with high scores overall. As evident in \autoref{tab:sb_performance}, across the six SBs in the performance tests, scores ranged from 20 to 21 out of a maximum total of 21. Four of the six LLM-generated SBs, the S-band observations of GW170817, the S-band observation of SN\,2025ulz, C-band observation of SN\,2026gzf, and the X-band observation of SN\,2024rjw passed all 21 rubric criteria. The remaining SBs, the Ku-band observation of SN\,2025ulz, and the multi-band observation of SN\,2021bmf both received a score of 20 for failing the \texttt{Total\_duration} criterion. These SBs both had a total SB duration that was longer than the prompted time constraint. Thus across the four remaining LLM-generated SBs, 83 of the 84 rubric criteria were passed, corresponding to a rule-level pass fraction of 98.8\%. We stress that two of the LLM-generated SBs that were used in training and marked as development targets (SN\,2024rjw and SN\,2021bmf), are not counted in the total passing score and are kept as consistency checks only. 

Discrepancies in total SB duration between the ground-truth and LLM-generated SBs are listed in \autoref{tab:sb_times_rms}. They arise from two causes. The first cause appears to be the presence of reference pointing scans that are recommended for observing frequencies $\gtrsim 10$\,GHz (Ku-band and above). In cases when those scans are present, the LLM seems to correctly enforce the VLA-recommended minimum of 2\,min\,30\,s on-source for the reference pointing, but fails to reduce the time duration of other scans (the most obvious being the slew to the phase calibrator scan) to adapt for that requirement and still fit within the total SB time requested by the user. This choice affected the two SBs that were scored 20/21 by \texttt{SCORE}. SN\,2021bmf multi-band shows a  $20$\,s overrun: the LLM combined the 3\,min\,50\,s slew scan seen in the development SBs with the VLA-recommended minimum of 2\,min\,30\,s on-source for reference pointing, yielding a single 6\,min\,20\,s scan that resulted in the total SB duration exceeding the requested one by 20\,s. SN\,2025ulz Ku-band shows a 3\,min\,50\,s total overrun in \autoref{tab:sb_times_rms}, as it suffered from both issues: 3\,min from the repeated \texttt{LOOP-START} scan, and an extra 50\,s from enforcing the reference scan duration but failing to compensate by shortening the duration of e.g. the slew scan (\autoref{app:AT2025ulz_KuSB}). 

The second cause of discrepancies in total SB duration between the ground-truth and LLM-generated SBs is the \texttt{LOOP-START} parameter in the phase-calibrator/target observing sequence. Following best practices, target scans need to be ``sandwiched'' between phase-calibrator scans.  The \texttt{LOOP-START} parameter, when set to ``Y'' (highlighted in red in \autoref{lst:GW_1min_discrepent}), instructs the OPT to repeat the first scan of the loop (i.e., the phase calibrator scan) once more after the final loop iteration. Because the LLM-generated SBs already contain an out-of-loop phase-calibrator scan that is executed after all of the loop iterations are completed, the extra scan added by setting the \texttt{LOOP-START} to Y is redundant, and it results in extra time added to the total SB duration. More importantly, this extra time is not recognized by the LLM due to the implicit nature of the flag. Both GW170817 and SN\,2025ulz S-band received a 21/21 score yet show a 1\,min overrun for this reason: the LLM did not include the loop-closure scan in its calculation of the global SB time constraint, so the SBs received a perfect score from \texttt{SCORE}.

Overall, the above described outcome highlights both that the LLM needs to be explicitly instructed on the behavior of the \texttt{LOOP-START} parameter, and that \texttt{SCORE} should be modified to flag such occurrences. Both issues, however, are easy to correct.

Testing the generated SBs within the VLA's OPT provides an important second stage of validation, as the OPT can simulate an actual observation, thus accounting for factors such as slew times and setup transitions that can reduce the available on-source integration time. The last is key to ensuring that a sensitivity comparable to the human-generated SBs is reached in the observations. Therefore, we uploaded each LLM-generated and ground-truth SB into the VLA OPT and compared the target on-source integration times and hardware resources, while manually calculating the expected noise root-mean-square (RMS) sensitivity based on the OPT static reports. The results of these additional checks, reported in \autoref{tab:sb_times_rms}, show that the calculated RMS values are generally comparable between the LLM-generated and ground-truth SBs, even though discrepancies are present in the total on-source integration times. Differences in loop structure also resulted in different allocations of observing time between the target and calibrators. In both Ku-band tests, the OPT initially returned errors indicating that reference-pointing scans had no calculated on-source integration time. This issue was resolved by adjusting the LST start range, as well as the antenna-wrap selection. These are checks that can only be carried out in the OPT and require minimal effort by the observer.

\section{Summary and conclusion}
\label{sec:conclusion}

We have presented a second-generation implementation of \texttt{RADAR} that resolves three limitations identified in \citet{2025ApJS..280...71P}: the fidelity of automated metadata extraction from GCN Circulars, the computational cost of federated radio inference, and the lack of a mechanism for translating analysis results into observation scheduling blocks. 
\paragraph{Improved GCN parsing.} \texttt{GPT-5.5} achieves an event-match $F_1$ of $\approx 0.893$, a 16\% improvement over \texttt{GPT-4.1} ($\approx 0.770$), and leads on GCN-level recall with $R_{\mathrm{GCN}} \approx 0.794$, a 10\% improvement over \texttt{GPT-4.1}; Claude-Opus-4.7 attains the highest precision ($\approx 0.978$). These gains reflect model maturation, not prompt redesign (\autoref{tab:metrics}). 

\paragraph{Accelerated federated inference.} Profiling revealed that network round-trip overhead through the \texttt{Octopus} fabric, not afterglow model evaluation, dominated the 3--4~hr fitting latency of the original pipeline (\autoref{fig:perf-trace}). Server-side thread pools and site-side subprocess pools for concurrent likelihood dispatch recover ${\approx}\,120$~samples\,s$^{-1}$ at $n_{\rm walkers} = 1024$ and $n_{\rm t} \geq 64$, matching local parallel \texttt{emcee} and reducing the fit to $\approx\,6$~min on 16-core hardware (\autoref{fig:scaling}). The neural-network surrogate \texttt{FIESTA} \citep{fiestaem} is ${\approx}\,5\times$ faster than \texttt{afterglowpy} per evaluation, but Python GIL contention under thread-based parallelism currently prevents it from matching \texttt{afterglowpy}'s subprocess scaling (\autoref{fig:fiesta}); \texttt{JAX}-native vectorization via \texttt{jax.vmap} is the natural next step. 

\paragraph{LLM-driven VLA scheduling.}\label{sec:Agentic VLA scheduling} The LLM-generated SBs for  multi-band continuum observations successfully uploaded to the VLA OPT (\autoref{tab:sb_performance}). Hence, our results show that it is possible to automate SB production using LLM and that the LLM-generated SBs can achieve a high ($>95\%$) rule-level pass fraction, although care must be taken to instruct the LLM on valid SB structure and \texttt{LOOP-START} parameter. 

\paragraph{From components to an autonomous follow-up loop.} These three capabilities, rapid federated fitting, reliable metadata extraction, and automated observation planning, constitute the core stages of a closed-loop follow-up system. In the current \texttt{RADAR} architecture, a GW alert triggers AI classifiers at distributed detector endpoints; the GCN parser assembles a radio light curve from public circulars and, where available, proprietary data that never leave the originating site; the federated MCMC engine fits a structured-jet afterglow model to the combined dataset; and the scheduling agent drafts the next VLA observation. The full chain now operates on a timescale of minutes. Further development could be directed toward setting thresholds on the GW classification to gate the follow-up trigger, improve goodness-of-fit criteria on the afterglow model to decide whether additional data are needed, and allowing an automated upload of the SBs in the VLA OPT. A human checkpoint before submission to oversubscribed facilities is essential at this stage and is incorporated by design.

The scientific motivation for \texttt{RADAR} is quantitative. Current GW detector runs yield BNS candidates at a rate of order one per month~\citep{2020LRR....23....3A, 2022ApJ...924...54P}, each requiring weeks of radio monitoring to constrain jet structure and energetics \citep{Hallinan_2017,2018Natur.561..355M, 2021ARA&A..59..155M,2026arXiv260405128O}. Next-generation GW detectors such as Cosmic Explorer and the Einstein Telescope will raise this to ${\sim}\,10^{5}$~events per year \citep{2017CQGra..34d4001A, 2020JCAP...03..050M,2023arXiv230613745E}, and the challenge will shift from detection to coordinated response. \texttt{RADAR} is designed for that regime: its federated architecture protects data rights, its modular design allows independent component upgrades, and the work presented here demonstrates that each stage already operates at latencies compatible with radio afterglow evolution. Extending the framework to additional facilities and wavelengths, deploying on the live alert stream, and implementing closed-loop scheduling with robust safeguards are the focus of our ongoing work.

\section*{Software and Data Availability}

The \texttt{RADAR} framework, including the federated MCMC server and site
implementations with the concurrent likelihood-evaluation modifications of
\autoref{sec:radioafterglow}, the \texttt{SCORE} scheduling-block
evaluator, and the benchmarking and profiling scripts used to produce
Figures~\ref{fig:perf-trace}--\ref{fig:fiesta}, is openly available at
\url{https://github.com/diaspora-project/radar-ii}.

The complete prompt sets are included in this article: the GCN-parser prompts
in \autoref{app:prompts} and the \texttt{SAGE} scheduling-block prompt
and reference-material list in \autoref{app:SB}. The raw LLM outputs
underlying \autoref{tab:metrics}, the six
LLM-generated scheduling blocks and their ground-truth counterparts, and the
\texttt{SCORE} rubric results of \autoref{tab:sb_performance} are also available
at \url{https://github.com/diaspora-project/radar-ii}.
The GCN Circulars analyzed in \autoref{sec:llm} are public
(\url{https://gcn.nasa.gov}), and the human-curated ground-truth catalog is
Table~3 of \citet{2025ApJS..280...71P}.

\software{\texttt{afterglowpy} \citep{Ryan_2020},
\texttt{emcee} \citep{foreman2013emcee},
\texttt{FIESTA} \citep{fiestaem},
\texttt{JAX} \citep{jax},
\texttt{Octopus} \citep{pan2024octopus}.}

\begin{acknowledgments}
We thank Hauke K\"{o}hn for helpful feedback regarding the use of  
\texttt{FIESTA}. A.C. and T.O. acknowledge support from the National Science Foundation (NSF) via grant AST-2431072 and from the U.S. Department of Energy (DoE) via grant DE-SC0025935. 
E.A.H. acknowledges support from NSF grants OAC-2514142 
and OAC-2209892. This work was partially supported by the US Department 
of Energy under contract No. DE-AC02-06CH11357, including funding 
from the Office of Advanced Scientific Computing Research (ASCR)'s 
Diaspora project and the Laboratory Directed Research and Development 
program. This research used resources of the Argonne Leadership 
Computing Facility, a U.S. Department of Energy Office 
of Science User Facility operated under contract DE-AC02-06CH11357. 
This research used both the DeltaAI advanced computing and data 
resource, which is supported by the NSF (award OAC 2320345) 
and the State of Illinois, and the Delta advanced computing 
and data resource, which is supported by the NSF (award OAC 2005572) 
and the State of Illinois. Delta and DeltaAI are joint efforts 
of the University of Illinois Urbana-Champaign and its NCSA.
\end{acknowledgments}

\clearpage

\appendix

\section{Model versions and decoding settings}
\label{app:llm-config}

\begin{table*}[h]
\centering
\caption{Model snapshots benchmarked in \autoref{sec:llm}. Limits and training cut-offs are as reported by the respective model providers.}
\label{tab:llm-config}
\begin{tabular}{llccc}
\hline\hline
Model & Provider & Context window (tokens) & Max.\ completion (tokens) & Training cut-off \\
\hline
\texttt{GPT-5.5}          & OpenAI    & 1,050,000 & 128,000 & 2025 December \\
\texttt{Claude-Opus-4.7}  & Anthropic & 1,000,000 & 128,000 & 2026 January \\
\texttt{Gemini-3.5-Flash} & Google    & 1,048,576 & 65,536  & 2025 January \\
\hline
\end{tabular}
\end{table*}

All three models were accessed through a common OpenAI-compatible endpoint, so identical code, prompts, and parameters were used throughout. Each prompt was sent as a single user message with no system prompt, no conversation history, and no tool use, so the only text available to the model was the Circular supplied in the prompt. Decoding used a sampling temperature of $1.0$ with \texttt{top\_p} at its default, decoding is therefore stochastic, which is why \autoref{tab:metrics} reports means and standard deviations over ten independent runs. The completion limit was $10{,}000$ tokens, far above any expected reply, and a reply that was not valid JSON or carried unexpected keys was reissued up to three times before the GCN Circular was recorded as unparsed.

\section{LLM prompts for GCN Parser}
\label{app:prompts}
For reproducibility we provide the prompt set of \citet{2025ApJS..280...71P} used to obtain the results of \autoref{sec:llm} and \autoref{tab:metrics}. The placeholder \texttt{<GCN CIRCULAR TEXT>} is replaced by the full plain-text body of the circular being parsed, and the model is instructed to answer with a strict JSON object so that the reply can be parsed programmatically without manual editing. Identical prompts were issued to all models.

The four prompts are applied to each circular in turn. Prompts~1 and~2 are always sent to LLMs. Prompt~3, which performs the actual measurement extraction, is issued only if Prompt~1 returns \texttt{true}. Prompt~4 is issued only when a regular-expression pre-filter finds more than one candidate R.A., Dec. in the circular, in which case the LLM is asked to select the pair that belongs to the optical transient; when exactly one candidate pair is found it is adopted directly, and when none is found the position fields are left empty. The non-detection flag returned by Prompt~2 overrides a detection claimed by Prompt~3 for the same circular.

\begin{lstlisting}[style=sbtext, caption={Prompt 1: Identification of circulars reporting radio follow-up of the GW170817 optical counterpart.}, label={lst:prompt1}]
Based on the individual GCN circular provided below, please decide if it does a radio follow-up on the optical transient counterpart to 
GW170817. This should be targeted at SSS17a. Please record "false" if the followup is not done on the optical transient counterpart and 
"true" if it is done on the optical transient counterpart (SSS17a)
The text of the GCN is as follows:
==================================
<GCN CIRCULAR TEXT>
==================================
Please make your response as a strict JSON formatted string without any additional text. The JSON should contain a single key-value pair
as either:
{"optical_transient": "true"}
or
{"optical_transient": "false"}
\end{lstlisting}

\begin{lstlisting}[style=sbtext, caption={Prompt 2: Identification of explicit non-detection statements.}, label={lst:prompt2}]
Based on the individual GCN circular provided below, please decide if it contains any statement claiming there is no significant emission
detected from the optical transient (called SSS17a). Please record "true" if the GCN includes a non-detection statement pertaining to the
transient. Please record "false" if the GCN does not include a non-detection statement.
The text of the GCN is as follows:
==================================
<GCN CIRCULAR TEXT>
==================================
Please make your response as a strict JSON formatted string without any additional text. The JSON should contain a single key-value pair 
as either:
{"non_emission_statement": "true"}
or
{"non_emission_statement": "false"}
\end{lstlisting}

\begin{lstlisting}[style=sbtext, caption={Prompt 3: Extraction of the radio measurements (frequency, flux density or upper limit, target name, and position).}, label={lst:prompt3}]
Based on the individual GCN circular provided below, please help me extract data that is only pertaining to the optical transient 
(SSS17a). For observations of the optical transient, could you please extract the frequencies of observation ("frequency"), any fluxes 
that were recorded ("flux_density"), and whether these are observations or upper limits ("type"). Please include units whenever possible. 
I would like to have the frequency in GHz and the flux density in mJy. Remember, 1 GHz = 1000000000 Hz. I also want you to record the 
name of the target which is being reported on ("name"), if it is given. If the target is not named, please do not report a name. Please 
also record the right ascension ("right_ascension") and declination ("declination"), if they are given. If not, please do not report them

Please use these conversions to report the flux density in mJy:

1 uJy = 1 microJy = 1 muJy
1000 uJy = 1 mJy = 1 milliJy
1000 mJy = 1 Jy

We do not count emission from NGC 4993, which is the host galaxy.
Please only discuss data available in this GCN, not other GCNs.

The text of the GCN is as follows:
==================================
<GCN CIRCULAR TEXT>
==================================

Please make your response as a strict JSON formatted string without any additional text. The JSON should contain any key-value pairs that 
are relevant to the optical transient (SSS17a). For example:
{
    "frequency": ...,
    "flux_density": ...,
    "name": ...,
    "right_ascension": ...,
    "declination": ...,
    "type": "upper_limit" | "observation"
}.

The expected keys are: "frequency", "flux_density", "name", "right_ascension", "declination", and "type". Please make sure accurate units 
are provided for frequency flux_density. If certain information is not available, please omit those keys from the JSON response. Please 
do not include any other information or context in the response. The JSON should only contain the relevant data extracted from the GCN.

Additional rules:
- If there are multiple observations (different frequencies, different times, OR different flux values at the same frequency), return one 
separate JSON object per observation. Never put multiple values into a single array-valued field, and never deduplicate observations that 
share the same frequency.
- Report frequencies exactly as written in the GCN. Do not round or approximate.
- Report flux density as a plain numeric value with unit only. Do not include < or ~ prefixes, and do not append suffixes like /beam or 
rms.
- Always include "frequency" and "flux_density" if at all possible. If the flux density value contains extra text or expressions, still 
report the numeric part in mJy and ignore the rest.
\end{lstlisting}

\begin{lstlisting}[style=sbtext, caption={Prompt 4: Disambiguation of the transient position when a circular quotes several coordinate pairs.}, label={lst:prompt4}]
Below shows the text for an individual GCN circular:
==================================
<GCN CIRCULAR TEXT>
==================================
There is an event detected in this GCN which is most closely described by a frequency of <FREQUENCY> GHz and flux density <FLUX DENSITY> 
mJy, and comes from the optical transient counterpart, called SSS17a. From the lists below, could you please help me select the right 
ascension and declination that best describe the optical transient location? Please do not comment on any other galaxies or the host 
galaxy.

Here is the right ascension list: <CANDIDATE RA LIST>
Here is the declination list: <CANDIDATE DEC LIST>

Please make your response as a strict JSON formatted string without any additional text. The JSON should contain two keys, 
"best_right_ascension" and "best_declination". For example:
{
    "best_right_ascension": ...,
    "best_declination": ...
}.
\end{lstlisting}

\section{VLA scheduling-block initial training, template and LLM prompt}
\label{app:SB}

\label{lst:LLM-SB direct training information links}
\begin{lstlisting}[style=sbtext, caption={
Collection of VLA documentation links and guidance provided to the
LLM as in-context reference for generating valid SBs; reproduced verbatim, including typos.}, label={lst:training}]
The list below provides use links with brief explination of how they will be usefully in creating a valid Schedule block (SB) for the 
VLA's Observation preperation tool (OPT):
==================================
Guide to Observing with the VLA Complete Manual - The complete observing manual for the VLA can be found here, specific sections are 
broken up and porivded below:
{https://science.nrao.edu/facilities/vla/docs/manuals/obsguide/referencemanual-all-pages}

OPT SB syntax and structure - This provides information on what a valid SB should contain and the proper syntax needed to import into the 
OPT:
{https://science.nrao.edu/facilities/vla/docs/manuals/opt-manual/text-files-and-catalogs-opt/importing-scan-lists}

1:VLA Frequency Bands and Samplers, 2:Frequency Bands and Tunability - Provides information of the acceptable frequency bands, samplers 
and possible restrictions:
1:{https://science.nrao.edu/facilities/vla/docs/manuals/propvla/frequency-bands-and-samplers}
2:{https://science.nrao.edu/facilities/vla/docs/manuals/oss/performance/vla-frequency-bands-and-tunability}

A STRICT list of acceptable hardware names, you cannot deviate form this list when choosing a hardware resource for scans: 
C-point, X-point, Q64f2A, Q64f3DCB, Ka64f2A, Ka64f3DCB, K64f3DCB, K64f2A, Ku48f3DCB, Ku48f2A, Ku48f3DCBalt, Ku48f2Aalt, Ku-slew, 
X16f3DCBalt, X16f2A, X16f3DCB, X32f3DCBalt, X32f2A, X32f3DCB, X16f2Aalt, X-slew, X32f2Aalt, C16f5DCalt, C32f3Bmixalt-blank, C16f5DC, 
C16f2A, C32f2Amixalt-blank, C32f5DC, C32f2A, C16f2Aalt, C32f3B, C16f3B, C16f3Balt, C32f5DCalt, C32f3Balt, C32f2Amixalt, C-slew, 
C32f3Bmixalt, C32f2Aalt, S16f2A, S16f3B, S16f5DC, S-slew, S14f2Ashiftalt-blank, S14f2Ashiftalt, S16f2Aalt, S16f5DCalt, S16f3Balt, L16f3B, 
L16f5DCalt, L16f2A, L16f3Balt, L16f2Aalt, L-slew, L16f5DC, 4P19f2DCBA, P16f2DCBA, P-slew, 4_3f2DCBA, eLWA (from semester 22A).

VLA array-configuration schedule - This provides the current and future configuration scheudle for the VLA:
{https://science.nrao.edu/facilities/vla/proposing/configpropdeadlines}

Flux Density Calibration - This provides information on the acceptable flux calibrators and when they can be used directly along with any 
special restrictions:
{https://science.nrao.edu/facilities/vla/docs/manuals/cal/flux}

List of VLA Calibrators - An list of acceptable VLA calibrators including the VLA calibrator structure/quality code that labels which 
calibraotrs would be compact and make suitable for calibration for the respective band and configuration:
{https://science.nrao.edu/facilities/vla/docs/observing/callist}

High Frequency Strategy - describes High-frequency refrence pointing that should be used for Ku, K, Ka, and Q-bands:
{https://science.nrao.edu/facilities/vla/docs/manuals/obsguide/topical-guides/hifreq}

Setup scans, 8-bit/3-bit, and requantizer scans - Overview of setup scans and examples of both 8-bit and 3-bit scans:
{https://science.nrao.edu/facilities/vla/docs/manuals/obsguide/set-up}

Phase-target calibration and cycle times - Calibration basics including when, how, and calibration recomendations as well as calibration 
phase-target cycle times:
{https://science.nrao.edu/facilities/vla/docs/manuals/obsguide/calibration}

Sensitivity and on-source time - When creating a SB we want to reach the lowest sensitivity possible within the prompted time constraint. 
You should aim to maximize target on-source time while preserving valid calibration structure, sensitivity information can be found here:
{https://science.nrao.edu/facilities/vla/docs/manuals/oss2017A/performance/sensitivity}

SCT: Importing Source Lists - When prompted by a user to observe a specific target a seperate SCT.txt file will need to be created, this 
will hold the name, RA and DEC of the desired source. The user will provide you the name of the source and its coordinate position, the 
syntax that needs to be followed for creating a importable SCT.txt file can be found here:
{https://science.nrao.edu/facilities/vla/docs/manuals/opt-manual/text-files-and-catalogs-opt/source-lists}

\end{lstlisting}
We note that while the LLM can follow the VLA's directions for loop-structure, the LLM cannot run the \href{https://obs.vla.nrao.edu/ect/}{VLA exposure calculator}. After importing an LLM-generated SB, the total on-source time should be manually checked with the exposure calculator for a desired observation sensitivity. 

Windows of observation should also be independently checked by the observer for possible RFI due to sun and moon distances with the \href{https://www.vla.nrao.edu/astro/guides/suncheck/}{VLA Sun and Moon Distance Checking Tool}.

When generating any SB, the LLM's choice of LST ranges were made by using a combination of the target R.A., SB duration, source declination, and calibrator positions. Its logic was to chose an LST start such that the target would transit near the midpoint of the SB. The LST windows chosen by the LLM should be treated as initial scheduling estimates where the final validation is conducted by importing the generated SB into the VLA's OPT.

\label{lst:souce catalog}
\begin{lstlisting}[style=sbtext, caption={Example of a importable SCT.txt file that contains more than one source, a single source SCT.txt file can also be used by only listing one source under the commented `\#' lines. This syntax is used to create a catalog of sources that are called upon by the generated SBs.}, label={lst:SRC_CAT_skeleton}]
* test1
# VLA SCT/PST source-list format
# Science targets used by the AI-test scheduling-block tests.
# Coordinates are Equatorial J2000. Velocity information is intentionally blank.

SN2021bmf;;Equatorial;J2000;16:33:29.416;-06:22:49.5;;;;N;
SN2024rjw;;Equatorial;J2000;21:03:10.107;+20:45:07.58;;;;N;
SN2025ulz;;Equatorial;J2000;15:51:54.201;+30:54:08.67;;;;N;
SN2026gzf;;Equatorial;J2000;09:59:42.889;+00:25:06.38;;;;N;
GW170817;;Equatorial;J2000;13:09:48.0850;-23:22:53.343;;;;N;
\end{lstlisting}
When importing a source.txt into the Source Catalog Tool (SCT) choose PST as the import format. Once uploaded, a source catalog can only be adjusted manually; if you want to add to the list of sources you will need to either delete the current source catalog and re-upload the new list of sources in the source.txt file or upload a new source.txt with the new targets declared.

\begin{lstlisting}[style=sbtext, caption={LLM-Prompt: Example of generating a valid SB in specified configuration band and time frame.}, label={lst:LLM-SB prompt}]
Using the proper syntax provided from the VLA SCT: Importing Source Lists, create ProgramName.txt file for source SN2020jqm, 
RA:13h 49m 18.564s DEC:-3d 46' 8.27", the name of the source catalog should be "test1".

Using the provided VLA documents covering the necessary checks and validations for the VLA's OPT schedule block imports, generate a 
seperate VLA Schedule block .txt file in A-configuration, C-band, targeting source SN2020jqm. When selecting hardware 
resources ensure that the receiver choice matches the telescope's configuration. The SB should have a total time of strictly 1 hour, 
attempt to reach the lowest sensitivity possible within this time. Add "test1" to the SRC-CAT. Report in the comment section of the SB 
header the phase-cal, flux-cal names used, the target name, RA and DEC, total SB duration and hardware resource used. Inform me of any 
information you were unable to find and or assumptions that were made.
\end{lstlisting}

\label{app:sb_skeleton}
\begin{lstlisting}[style=sbtext, caption={Shortened VLA OPT scheduling-block.txt file structure used for the initial SN2024rjw continuum test.}, label={lst:sb_skeleton}]
VERSION; 7;

 SRC-CAT; test1, VLA;
HDWR-CAT; NRAO Defaults;

SCHED-BLOCK; SN2024rjw_Cband_test; Dynamic; ...;

STD; Dummy; 1331+305=3C286; C32f2Amixalt-blank; DUR; ...; SetAtnGain,; ;

LOOP-START; FluxCal-3C286; ...;
STD; ; 1331+305=3C286; C32f2Amixalt-blank; DUR; ...; CalBP,CalFlux,; ;
LOOP-END;

STD; Slew2PhaseCal; J2035+1857; C32f2Amixalt-blank; DUR; ...; CalGain,; ;

LOOP-START; PhaseCal2SourceLoop; ...;
STD; ; J2035+1857; C32f2Amixalt-blank; DUR; ...; CalGain,; ;
STD; ; SN2024rjw; C32f2Amixalt-blank; DUR; ...; ObsTgt,; ;
STD; ; SN2024rjw; C32f2Amixalt-blank; DUR; ...; ObsTgt,; ;
LOOP-END;
\end{lstlisting}

\label{app:GW_1min_discrepent}
\begin{lstlisting}[style=sbtext, escapeinside={(*@}{@*)}, caption={LLM-generated scheduling-block.txt file of GW170817 showing a 1\,min overrun, the red parameters show the repeated scan.}, label={lst:GW_1min_discrepent}]
VERSION; 7;

 SRC-CAT; AI-test, VLA;
HDWR-CAT; NRAO Defaults;

SCHED-BLOCK; GW170817_Sband_Bconfig_3h30m_AItest; Dynamic; 1; 2027-03-03 00:00:00,2027-06-14 23:59:00; 11:00:00-15:30:01, ; 0.0; B; 445.0; 20.0; N; N; S; GW170817 deep continuum imaging B configuration S band target RA 13h09m48.0850s Dec -23d22m53.343s phase calibrator 1258-223 flux bandpass delay calibrator 3C286 total duration approximately 3h30m on-source time approximately 155m resource S16f3B; 

  STD; S-Slew; 1331+305=3C286; S-slew; DUR; 0h 8m 0s; CCW; N; N; N; N; Y; N; N; SetAtnGain,; ;
  STD; S-Attn; 1331+305=3C286; S16f3B; DUR; 0h 1m 0s; CCW; N; N; N; N; Y; N; N; SetAtnGain,; ;
  STD; S-Req; 1331+305=3C286; S16f3B; DUR; 0h 0m 10s; CCW; N; N; N; N; Y; N; N; SetAtnGain,; ;
  LOOP-START; S-FluxCal-3C286; 9; N; ;
    STD; S-FluxCal=3C286; 1331+305=3C286; S16f3B; DUR; 0h 1m 0s; ; N; N; N; N; Y; N; N; CalBP,CalFlux,CalDelay,; ;
  LOOP-END;
  STD; S-slew2PhaseCal; 1258-223; S16f3B; DUR; 0h 3m 50s; ; N; N; N; N; Y; N; N; CalGain,; ;
  LOOP-START; S-PhaseCal2Source; 31; (*@\textcolor{red}{\textbf{Y}}@*); ;
    STD; ; 1258-223; S16f3B; DUR; 0h 1m 0s; ; N; N; N; N; Y; N; N; CalGain,; ;
    STD; ; GW170817_7as; S16f3B; DUR; 0h 1m 0s; ; N; N; N; N; Y; N; N; ObsTgt,; ;
    STD; ; GW170817_7as; S16f3B; DUR; 0h 1m 0s; ; N; N; N; N; Y; N; N; ObsTgt,; ;
    STD; ; GW170817_7as; S16f3B; DUR; 0h 1m 0s; ; N; N; N; N; Y; N; N; ObsTgt,; ;
    STD; ; GW170817_7as; S16f3B; DUR; 0h 1m 0s; ; N; N; N; N; Y; N; N; ObsTgt,; ;
    STD; ; GW170817_7as; S16f3B; DUR; 0h 1m 0s; ; N; N; N; N; Y; N; N; ObsTgt,; ;
  LOOP-END;
  STD; S-FinalPhaseCal; 1258-223; S16f3B; DUR; 0h 2m 0s; ; N; N; N; N; Y; N; N; CalGain,; ;
\end{lstlisting}

\begin{lstlisting}[style=sbtext, escapeinside={(*@}{@*)}, caption={LLM-generated scheduling-block.txt file of SN2025ulz Ku-band showing both the 3\,min overrun due to loop strucure and 50 seconds added by chosen slew time.}, label={app:AT2025ulz_KuSB}]
VERSION; 7;

 SRC-CAT; AI-test, VLA;
HDWR-CAT; NRAO Defaults;

SCHED-BLOCK; SN2025ulz_Kuband_Dconfig_3h25m_Ku48f3DCB_AItest; Dynamic; 1; 2026-07-10 00:00:00,2026-10-19 23:59:00; 09:05:00-11:15:01, ; 0.0; D; 445.0; 35.0; N; N; Ku; SN2025ulz deep continuum imaging D configuration Ku band target RA 15h51m54.201s Dec +30d54m08.67s phase calibrator J1602+3326 also used for X-band reference pointing flux bandpass delay calibrator 3C286 total duration approximately 3h25m50s, target on-source time approximately 142.5m, resource Ku48f3DCB, repeated reference pointing included about hourly; 

  STD; Ku-XPointSetup; 1331+305=3C286; X-point; DUR; 0h 6m 0s; CCW; N; N; N; N; Y; N; N; SetAtnGain,; ;
  STD; Ku-Attn; 1331+305=3C286; Ku48f3DCB; DUR; 0h 1m 0s; CCW; N; N; N; N; Y; N; N; SetAtnGain,; ;
  PTG; Ku-RefPoint-3C286; 1331+305=3C286; X-point; DUR; 0h 7m 0s; CCW; N; N; N; N; Y; ;
  STD; Ku-Req; 1331+305=3C286; Ku48f3DCB; DUR; 0h 0m 30s; CCW; Y; N; N; N; Y; N; N; SetAtnGain,; ;
  LOOP-START; Ku-FluxCal-3C286; 3; N; ;
    STD; Ku-FluxCal=3C286; 1331+305=3C286; Ku48f3DCB; DUR; 0h 1m 0s; CCW; Y; N; N; N; Y; N; N; CalBP,CalFlux,CalDelay,; ;
  LOOP-END;
  STD; Ku-slew2PointCal; J1602+3326; Ku48f3DCB; DUR; (*@\textcolor{red}{\textbf{0h 3m 50s}}@*); ; Y; N; N; N; Y; N; N; CalGain,; ;
  PTG; Ku-RefPoint-NearTarget-1; J1602+3326; X-point; DUR; 0h 3m 30s; ; N; N; N; N; Y; ;
  STD; Ku-Req-NearTarget-1; J1602+3326; Ku48f3DCB; DUR; 0h 0m 30s; ; Y; N; N; N; Y; N; N; SetAtnGain,; ;
  LOOP-START; Ku-PhaseCal2Source-1; 10; (*@\textcolor{red}{\textbf{Y}}@*); ;
    STD; Ku-PhaseCal; J1602+3326; Ku48f3DCB; DUR; 0h 1m 0s; ; Y; N; N; N; Y; N; N; CalGain,; ;
    STD; SN2025ulz_Ku_1as; AT2025ulz; Ku48f3DCB; DUR; 0h 1m 0s; ; Y; N; N; N; Y; N; N; ObsTgt,; ;
    STD; SN2025ulz_Ku_1as; AT2025ulz; Ku48f3DCB; DUR; 0h 1m 0s; ; Y; N; N; N; Y; N; N; ObsTgt,; ;
    STD; SN2025ulz_Ku_1as; AT2025ulz; Ku48f3DCB; DUR; 0h 1m 0s; ; Y; N; N; N; Y; N; N; ObsTgt,; ;
    STD; SN2025ulz_Ku_1as; AT2025ulz; Ku48f3DCB; DUR; 0h 1m 0s; ; Y; N; N; N; Y; N; N; ObsTgt,; ;
    STD; SN2025ulz_Ku_1as; AT2025ulz; Ku48f3DCB; DUR; 0h 0m 45s; ; Y; N; N; N; Y; N; N; ObsTgt,; ;
  LOOP-END;
  PTG; Ku-RefPoint-NearTarget-2; J1602+3326; X-point; DUR; 0h 3m 30s; ; N; N; N; N; Y; ;
  STD; Ku-Req-NearTarget-2; J1602+3326; Ku48f3DCB; DUR; 0h 0m 30s; ; Y; N; N; N; Y; N; N; SetAtnGain,; ;
  LOOP-START; Ku-PhaseCal2Source-2; 10; (*@\textcolor{red}{\textbf{Y}}@*); ;
    STD; Ku-PhaseCal; J1602+3326; Ku48f3DCB; DUR; 0h 1m 0s; ; Y; N; N; N; Y; N; N; CalGain,; ;
    STD; SN2025ulz_Ku_1as; AT2025ulz; Ku48f3DCB; DUR; 0h 1m 0s; ; Y; N; N; N; Y; N; N; ObsTgt,; ;
    STD; SN2025ulz_Ku_1as; AT2025ulz; Ku48f3DCB; DUR; 0h 1m 0s; ; Y; N; N; N; Y; N; N; ObsTgt,; ;
    STD; SN2025ulz_Ku_1as; AT2025ulz; Ku48f3DCB; DUR; 0h 1m 0s; ; Y; N; N; N; Y; N; N; ObsTgt,; ;
    STD; SN2025ulz_Ku_1as; AT2025ulz; Ku48f3DCB; DUR; 0h 1m 0s; ; Y; N; N; N; Y; N; N; ObsTgt,; ;
    STD; SN2025ulz_Ku_1as; AT2025ulz; Ku48f3DCB; DUR; 0h 0m 45s; ; Y; N; N; N; Y; N; N; ObsTgt,; ;
  LOOP-END;
  PTG; Ku-RefPoint-NearTarget-3; J1602+3326; X-point; DUR; 0h 3m 30s; ; N; N; N; N; Y; ;
  STD; Ku-Req-NearTarget-3; J1602+3326; Ku48f3DCB; DUR; 0h 0m 30s; ; Y; N; N; N; Y; N; N; SetAtnGain,; ;
  LOOP-START; Ku-PhaseCal2Source-3; 10; (*@\textcolor{red}{\textbf{Y}}@*); ;
    STD; Ku-PhaseCal; J1602+3326; Ku48f3DCB; DUR; 0h 1m 0s; ; Y; N; N; N; Y; N; N; CalGain,; ;
    STD; SN2025ulz_Ku_1as; AT2025ulz; Ku48f3DCB; DUR; 0h 1m 0s; ; Y; N; N; N; Y; N; N; ObsTgt,; ;
    STD; SN2025ulz_Ku_1as; AT2025ulz; Ku48f3DCB; DUR; 0h 1m 0s; ; Y; N; N; N; Y; N; N; ObsTgt,; ;
    STD; SN2025ulz_Ku_1as; AT2025ulz; Ku48f3DCB; DUR; 0h 1m 0s; ; Y; N; N; N; Y; N; N; ObsTgt,; ;
    STD; SN2025ulz_Ku_1as; AT2025ulz; Ku48f3DCB; DUR; 0h 1m 0s; ; Y; N; N; N; Y; N; N; ObsTgt,; ;
    STD; SN2025ulz_Ku_1as; AT2025ulz; Ku48f3DCB; DUR; 0h 0m 45s; ; Y; N; N; N; Y; N; N; ObsTgt,; ;
  LOOP-END;

\end{lstlisting}

\section{SB scoring criteria}

\begin{table*}[!ht]
\begin{center}
\caption{Scoring criteria used to evaluate LLM-generated VLA scheduling blocks. See \autoref{sec:score} for discussion.
\label{tab:scoring}}
\begin{tabular}{ccc}
\hline
{Rubric criterion} &
\colhead{Check type} &
\colhead{Pass criterion}\\
\hline
\texttt{version\_present} &
Mechanical &
\texttt{VERSION; 7;} is the first non-blank line \\
\texttt{src\_cat\_present} &
Mechanical &
A \texttt{SRC-CAT} line is present \\
\texttt{hdwr\_cat\_present} &
Mechanical &
An \texttt{HDWR-CAT} line is present \\
\texttt{loop\_closure} &
Mechanical &
Every \texttt{LOOP-START} has a matching \texttt{LOOP-END} \\
\texttt{dummy\_setup\_first} &
Mechanical &
The first scan is a dummy/setup scan with \texttt{SetAtnGain} intent \\
\texttt{hardware\_in\_allowlist} & Mechanical &
Every hardware resource matches a valid catalog resource name \\
\texttt{intents\_valid} &
Mechanical &
Every scan intent belongs to the allowed set \\
\texttt{flux\_present} &
Mechanical &
At least one \texttt{CalFlux} scan is present \\
\texttt{bandpass\_present} &
Mechanical &
At least one \texttt{CalBP} scan is present \\
\texttt{flux\_cal\_standard} &
Mechanical &
Flux calibrator is 3C286, 3C48, or 3C147 \\
\texttt{phase\_present} &
Mechanical &
At least one \texttt{CalGain} scan is present \\
\texttt{target\_present} &
Mechanical &
At least one \texttt{ObsTgt} scan is present \\
\texttt{one\_phase\_per\_target\_loop} &
Mechanical &
Each target loop contains exactly one phase-calibrator line \\
\texttt{template\_order} &
Mechanical &
Scan order follows setup $\rightarrow$ flux/bandpass $\rightarrow$ phase $\rightarrow$ target \\
\texttt{config\_matches\_date} &
Mechanical &
Array configuration matches the observing-date window \\
\texttt{highfreq\_refpointing} &
Mechanical &
Ku/K/Ka/Q-band SB contains X-band reference pointing or an explicit waiver \\
\texttt{cycle\_time} &
Advisory &
Phase-target cycle time is within NRAO guidance for the band/configuration \\
\texttt{Total\_duration} &
Computed &
Total duration of the SB does not exceed the prompt constraint \\
\texttt{coord\_SCT\_match} &
Computed &
Target coordinates in the generated SCT.txt file agrees with prompt target RA and DEC\\
\texttt{phase\_cal\_valid} &
External &
Phase calibrator for request band and configuration has code marked either P or S\\
\texttt{phase\_cal\_separation} &
External &
Phase-calibrator angular separation is within the band-dependent threshold \\
\hline\hline
\end{tabular}
\end{center}
\end{table*}

\bibliography{main}{}
\bibliographystyle{aasjournal}

\end{document}